\documentclass[preprint,showpacs,preprintnumbers,amsmath,amssymb]{revtex4}

\usepackage{graphicx}
\usepackage{dcolumn}
\usepackage{bm}

\begin{document}


\title{The Hall conductance, topological quantum phase transition and
the Diophantine equation on honeycomb lattice}

\author{Masatoshi Sato}
\email{msato@issp.u-tokyo.ac.jp}
\author{Daijiro Tobe}
\author{Mahito Kohmoto}%
\affiliation{%
The Institute for Solid State Physics, The University of Tokyo,
Kashiwanoha 5-1-5, Kashiwa, Chiba, 277-8581, Japan 
}%

\date{\today}

\begin{abstract}
We consider a tight-binding model with the nearest neighbour hopping
 integrals on
the honeycomb lattice in a magnetic field.
Assuming one of the three hopping integrals, which we denote $t_a$, can
 take a different value from the two others,    
we study quantum phase structures controlled by the anisotropy of the
 honeycomb lattice.
For weak and strong $t_a$ regions, respectively, 
the Hall conductances are calculated
 algebraically by using the Diophantine equation. 
Except for a few specific gaps, we completely determine the Hall conductances
 in these two regions
 including those for subband gaps.
In a weak magnetic field, it is found that the
 weak $t_a$ region shows the unconventional quantization of the Hall conductance,
 $\sigma_{xy}=-(e^2/h)(2n+1)$, $(n=0,\pm 1,\pm 2,\cdots)$, near the
 half-filling, while the strong $t_a$ region shows only the
 conventional one,  $\sigma_{xy}=-(e^2/h)n$, $(n=0,1,2,\cdots)$.
From topological nature of the Hall conductance, the existence of
gap closing points and quantum phase transitions in the intermediate
 $t_a$ region are concluded.
We also study numerically the quantum phase structure in
 detail, and find that even when $t_a=1$, namely in graphene case, the
 system is in the weak $t_a$ phase except when the Fermi energy is
 located near the van Hove singularity or the lower and upper edges of
 the spectrum.  
\end{abstract}

\pacs{Valid PACS appear here}
\maketitle

\section{Introduction}
\label{sec:introduction}

The purpose of this paper is to present some arguments for quantum Hall
conductivity on the honeycomb lattice.
Recently, quantized Hall effect is observed in graphene, and 
the Hall conductivity has been determined to be unconventionally
quantized \cite{NMMFKZJSG05,ZTSK05}, 
\begin{eqnarray}
\sigma_{xy}=-2(2n+1)\frac{e^2}{h}. 
\label{eq:uqhc}
\end{eqnarray}
(The factor $2$ comes from the spin degrees of freedom.) 
This unusual quantization is relevant when single particle physics
dominates the behavior of the system \footnote{Recent experimental
studies revealed new quantum Hall states which is not included in
(\ref{eq:uqhc}) \cite{ZJSPTFCJSK06, JZSK07}. The electron-electron
interaction presumably plays a
crucial role for them.}, and it has been considered as a
consequence of the existence of zero modes in graphene
\cite{ZA02,GS05,PGC06,NMMFKZJSG06}. 
In graphene, there exist zero modes in the absence of a magnetic
field due to the honeycomb lattice structure, and 
the unconventional quantum Hall conductance (\ref{eq:uqhc}) was explained
by treating these zero modes as Dirac fermions \cite{ZA02,GS05}.
However, it is not trivial whether the Dirac fermion argument is true or
not, because the Hall conductivity is given by an integral over whole
Brillouin zone while the Dirac-fermion argument looks only the zero mode
near the Fermi surface. 
Actually, it has been known that the logic of quantum Hall conductivity
using the Dirac fermions is not correct in general \cite{Oshikawa94}.
Furthermore, numerical analyses for the tight-binding model of graphene
have shown that the unconventional quantization persists up to the van
Hove singularity where the Dirac fermion argument is no longer valid
\cite{HFA06}. 

In this paper, we present an alternative explanation of the unconventional
quantization by using the tight-binding model on the honeycomb lattice.
For the tight-binding model on the honeycomb lattice, the energy
spectrum in a magnetic field was studied in \cite{Rammal85}, and
some numerical analyses for the Hall conductance have been done previously
\cite{HK06,HFA06,BHZCW06}.
However no analytical study of the Hall conductance was presented
systematically. 
Here we obtain the algebraic expression of the Hall conductances for
almost all gaps including subband gaps, and
explain why the unconventional quantum Hall effect (\ref{eq:uqhc})
persists up to the van Hove singularity.

To obtain the algebraic expression of the Hall conductances, 
we study effects of anisotropy of the hopping parameters on the
honeycomb lattice \cite{HKNK06,HK06, DPG08, MacDonald84}. 
For the square lattice, as suggested by Aubry-Andr\'{e}
duality\cite{AA80}, it is known that none of gaps closes when we change
the ratio of the hopping parameters, $t_x/t_y$, and this facilitates the
calculation of its Hall conductances \cite{TKNN82}.
For the honeycomb lattice, however, it will be found that gap closing
points appear since there is no duality.
We present here a detailed study of the (topological) phase structures
induced by the anisotropy of the hopping parameters on the honeycomb lattice, 
and determine its Hall conductances.

In the following, we describe the quantum Hall effect in terms of 
the Diophantine equation.
For later convenience we present a brief derivation of the relation
between the Hall conductance and the Diophantine equation \cite{Kohmoto92}.
First we use the St\v{r}eda formula \cite{Streda82,Streda82-2} for the
Hall conductance,
\begin{eqnarray}
\sigma_{xy}=-e\frac{\partial \rho}{\partial B}, 
\label{eq:streda}
\end{eqnarray}
where $\rho$ is the electron density and the derivative is to be taken
with the Fermi level fixed inside the gap.
Then at the same time, the Hall conductance is written as
\cite{TKNN82,Kohmoto85,ASS83}
\begin{eqnarray}
\sigma_{xy}=-\frac{e^2}{h}t_r,
\label{eq:TKNN}
\end{eqnarray}
when the Fermi level is in the $r$-th gap from the bottom, where
$t_r$ is guaranteed to be an integer given by the Chern number. 
Combining them, we have
\begin{eqnarray}
\frac{\partial \rho}{\partial B}=\frac{e}{h}t_r. 
\end{eqnarray}
The energy gaps are stable under small perturbation and therefore persist
under slight variation of $B$. Thus we obtain
\begin{eqnarray}
\rho=\frac{\mbox{\rm const.}}{v_0}+\frac{e}{h}B t_r, 
\label{eq:rho}
\end{eqnarray}
where $v_0$ is the area of a unit cell. (Here $\mbox{\rm const.}/v_0$ is a
constant of integration.)
When the flux per unit cell is $\Phi= p/q$ (in units of
$h/e$) where $p$ and $q$ are co-prime integers and $q>0$, the area of
the magnetic Brillouin zone is $((2\pi)^2/v_0 q)$, and the density of
electrons in a single band is given by $(1/v_0 q)$. 
Thus, when there are $r$ bands below the Fermi energy, the density of
electrons is
\begin{eqnarray}
\rho=\frac{r}{v_0 q}. 
\end{eqnarray}
Then (\ref{eq:rho}) is rewritten as 
\begin{eqnarray}
r=\mbox{\rm const.}\times q+pt_r. 
\end{eqnarray}
In this equation, $r$, $p$ and $t_r$ are integers, thus $\mbox{\rm
const.}\times q$ must be an integer. 
However, since $\mbox{\rm const.}$ is independent of $q$ and $q$ can change
when the magnetic field is varied without making a point of contact, 
then $\mbox{\rm const.}$ itself must be an integer $s_r$. 
Thus we have
\begin{eqnarray}
r=q s_r + p t_r, 
\end{eqnarray}
which is the Diophantine equation.

The plan of the remainder of the paper is as follows.
In Sec.\ref{sec:model}, we explain the tight-binding model on the
honeycomb lattice briefly. An anisotropic hopping parameter $t_a$ is
introduced in the tight-binding model.
In Sec.\ref{sec:WTP}, the weak $t_a$ limit is studied.
It is shown that the Hall conductances for almost all gaps including
subband ones are determined algebraically by
using the Diophantine equation.
Furthermore,  in a weak magnetic field, the
unconventional Hall effect (\ref{eq:uqhc}) is found to hold for visible
gaps. 
We study the strong $t_a$ limit in Sec.\ref{sec:STP}.
In this limit, our model on the honeycomb lattice is shown to reduce to
a pair of the tight-binding models on the square lattice.
Using this result, we establish an algebraic rule to determine the Hall
conductances in the strong $t_a$ limit.
It is found that the algebraic rule in the
strong $t_a$ limit is different from one in the weak $t_a$ limit. 
As a result, only conventional quantization of the Hall conductance is
obtained in this limit.
From the differences of the Hall conductances between the
two limits, the existence of the gap closing points and topological
quantum phase transitions in the intermediate $t_a$ region are concluded.
In Sec.\ref{sec:intermediate}, we study the topological quantum phase
transitions in the intermediate $t_a$ region in detail.
The existence of the gap closing points is confirmed numerically.
In addition, we will find that for some gaps near the van Hove
singularity at $t_a=1$ or on the lower and upper edges of the spectrum
unexpected topological quantum phase transitions occur due to
accumulation of gaps in a weak magnetic field.
In Sec.\ref{sec:graphene}, we apply our results to graphene where $t_a$
is given by $1$ and determine its Hall conductance.
The unconventional quantization of the Hall conductance is obtained from
our results. It is also explained naturally why it persists up to the
van Hove singularity.
In Sec.\ref{sec:graphene}, we also examine the edge states of graphene
in a weak magnetic field, in order to confirm our results using the
bulk-edge correspondence. 
Finally, in Sec.\ref{sec:conclusion}, we will summarize our results.

\section{Model}
\label{sec:model}

\begin{figure}[h]
\begin{center}
\includegraphics[width=6cm]{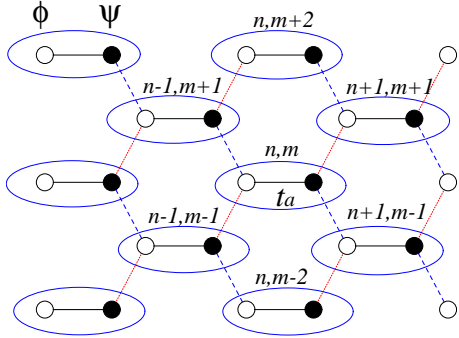}
\caption{The honeycomb lattice. Closed and open circles shows sublattice
 A and B, respectively. $t_a$ is the hopping integral of the horizontal bond.}
\label{fig:Honeylattice}
\end{center}
\end{figure}

Let us consider the tight-binding model on the honeycomb lattice with
the nearest neighbour hopping in a magnetic field.
See Fig.\ref{fig:Honeylattice}.
Denote wave functions on sublattices A and B as $\psi_{n,m}$ and
$\phi_{n,m}$, respectively, then the tight-binding model is given by 
\begin{eqnarray}
-\phi_{n+1,m-1}-\phi_{n+1,m+1}-t_ae^{-i\pi \Phi m}\phi_{n,m}+\xi\psi_{n,m}
=E\psi_{n,m},
\nonumber\\ 
-\psi_{n-1,m+1}-\psi_{n-1,m-1}-t_ae^{i\pi\Phi m}\psi_{n,m}-\xi\phi_{n,m}
=E\phi_{n,m},
\label{eq:tbm}
\end{eqnarray}
where a magnetic flux through a unit hexagon is given by $\Phi$,
the hopping integrals of the horizontal bonds are $t_a$ and those for the
other bonds are 1. 
Here we have introduced potentials $\xi$ on sublattice A and $-\xi$ on
sublattice B to remove a subtle singularity at $E=0$. 
We take $\xi\rightarrow 0$ in the final
stage of analysis. 
For simplicity we neglect the spin degrees of freedom in the following.

\section{Weak $t_a$ phase}
\label{sec:WTP}

In this section, we study the weak coupling limit of the horizontal bond
$t_a\ll 1$. 
Using the perturbation theory, we derive the Diophantine equation and
provide a general rule to determine the Hall conductances.
We also find that the unconventional
quantization (\ref{eq:uqhc}) holds in a weak magnetic field. 

\subsection{weak $t_a$ perturbation}
\label{subsec:WTP}
\begin{figure}[h]
\begin{center}
\includegraphics[width=6cm]{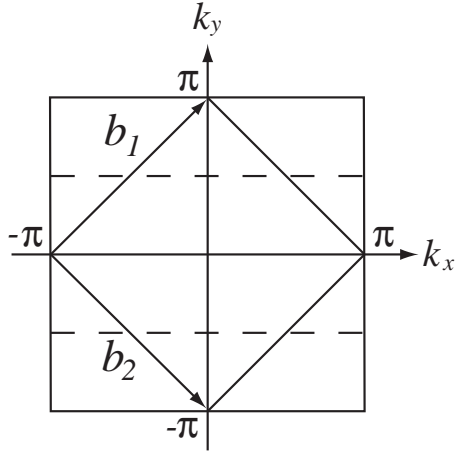}
\caption{The first and the second Brillouin zones for $\Phi=0$ in the momentum space
 defined by (\ref{eq:fourier}). 
The primitive vectors are ${\bm a_1}=(1,1)$ 
and ${\bm a_2}=(1,-1)$ in the space of $(n,m)$, so the generators of the
 reciprocal lattice 
 are ${\bm b}_1=\pi (1,1)$ and ${\bm b}_2=\pi (1,-1)$.}
\label{fig:bz}
\end{center}
\end{figure}

\begin{figure}[h]
\begin{center}
\includegraphics[width=6cm]{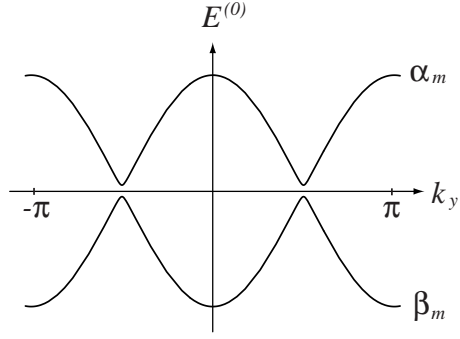}
\caption{The energy spectrum for $t_a=0$. The upper band and lower one
 correspond to the eigen state with $\alpha_m=1$ and one with
 $\beta_m=1$, respectively. We show the case of $\xi=0.05$. }
\label{fig:t=0}
\end{center}
\end{figure}

Let us consider (\ref{eq:tbm}) in the momentum space.
By performing the Fourier transformation
of $\psi_{n,m}$ and $\phi_{n,m}$,
\begin{eqnarray}
\psi_{n,m}=\sum_{k_x,k_y}e^{ik_x n+ik_y m}u({\bm k}),
\nonumber\\
\phi_{n,m}=\sum_{k_x,k_y}e^{ik_x n+ik_y m}v({\bm k}),
\label{eq:fourier}
\end{eqnarray}
(\ref{eq:tbm}) becomes
\begin{eqnarray}
h({\bm k})
\left[
\begin{array}{c}
u(k_x,k_y) \\
v(k_x,k_y)
\end{array}
\right]
-t_a \sigma_+
\left[
\begin{array}{c}
u(k_x,k_y+\pi\Phi) \\
v(k_x,k_y+\pi\Phi)
\end{array}
\right]
-t_a
\sigma_-
\left[
\begin{array}{c}
u(k_x,k_y-\pi\Phi) \\
v(k_x,k_y-\pi\Phi)
\end{array}
\right]
=E
\left[
\begin{array}{c}
u({\bm k}) \\
v({\bm k})
\end{array}
\right],
\label{eq:ftbm}
\end{eqnarray}
where $h({\bm k})$ and $\sigma_{\pm}$ are given by 
\begin{eqnarray}
h({\bm k})= \left[
\begin{array}{cc}
\xi &\Delta^{(0)}({\bm k}) \\
\Delta^{(0)*}({\bm k})& -\xi
\end{array}
\right],
\quad
\Delta^{(0)}({\bm k})=-2e^{ik_x}\cos k_y, 
\end{eqnarray}
and 
\begin{eqnarray}
\sigma_+=\left[
\begin{array}{cc}
0 &1 \\
0 &0
\end{array}
\right],
\quad
\sigma_-=\left[
\begin{array}{cc}
0 &0 \\
1 &0
\end{array}
\right].
\end{eqnarray}
When $t_a=0$, (\ref{eq:ftbm}) reduces to
\begin{eqnarray}
h({\bm k})
\left[
\begin{array}{c}
u^{(0)}({\bm k}) \\
v^{(0)}({\bm k})
\end{array}
\right]
=E
\left[
\begin{array}{c}
u^{(0)}({\bm k}) \\
v^{(0)}({\bm k})
\end{array}
\right],
\end{eqnarray}
then the energies $E$ are found to be
\begin{eqnarray}
E=\pm \sqrt{\xi^2+4\cos^2 k_y}
(\equiv\pm E^{(0)}(k_y)),
\end{eqnarray}
and the eigen vector with the positive energy, $E=E^{(0)}(k_y)$, is 
\begin{eqnarray}
\left[
\begin{array}{c}
u^{(0)}_+({\bm k})\\
v^{(0)}_+({\bm k})
\end{array}
\right]
=\frac{1}{\sqrt{2E^{(0)}(k_y)(\xi+E^{(0)}(k_y))}}
\left[
\begin{array}{c}
\xi+E^{(0)}(k_y) \\
\Delta^{(0)*}({\bm k})
\end{array}
\right],
\label{eq:u+0v+0}
\end{eqnarray}
and that with the negative one, $E=-E^{(0)}(k_y)$, is
\begin{eqnarray}
\left[
\begin{array}{c}
u^{(0)}_-({\bm k})\\
v^{(0)}_-({\bm k})
\end{array}
\right]
=\frac{1}{\sqrt{2E^{(0)}(k_y)(\xi+E^{(0)}(k_y))}}
\left[
\begin{array}{c}
-\Delta^{(0)}({\bm k}) \\
\xi+E^{(0)}(k_y)
\end{array}
\right].
\label{eq:u-0v-0}
\end{eqnarray}
Here the staggered potential $\xi$ removes singularities in the
energies, $\pm E^{(0)}(k_y)$, and the eigen vectors at
$k_y=\pm \pi /2$.
It also remedies the singularity of ${\bm A}({\bm k})$ in
(\ref{eq:gaugefield}) at $k_y=\pm \pi/2$. 
To solve (\ref{eq:ftbm}) in perturbative expansions in powers
of $t_a$, 
it is convenient to take the basis $(u({\bm k}), v({\bm k}))$ as
\begin{eqnarray}
\left[
\begin{array}{c}
u({\bm k}) \\
v({\bm k})
\end{array}
\right]
=
\alpha({\bm k})
\left[
\begin{array}{c}
u_+^{(0)}({\bm k}) \\
v_+^{(0)}({\bm k})
\end{array}
\right]
+\beta({\bm k})
\left[
\begin{array}{c}
u_-^{(0)}({\bm k}) \\
v_-^{(0)}({\bm k})
\end{array}
\right].
\end{eqnarray}
Substituting this into (\ref{eq:ftbm}) and multiplying $(u_+^{(0)*}({\bm
k}),v_+^{(0)*}({\bm k}))$ to the both sides, we obtain 
\begin{eqnarray}
&&\left(E-E^{(0)}(k_y^0+\pi\Phi m)\right)\alpha_m(k_x,k_y^0)=
\nonumber\\
&&-t_a
\left[
u_+^{(0)*}(k_x,k_y^0+\pi\Phi m)v_+^{(0)}(k_x,k_y^0+\pi\Phi(m+1)) 
\right]
\alpha_{m+1}(k_x,k_y^0)
\nonumber\\
&&-t_a
\left[
u_+^{(0)*}(k_x,k_y^0+\pi\Phi m)v_-^{(0)}(k_x,k_y^0+\pi\Phi(m+1)) 
\right]
\beta_{m+1}(k_x,k_y^0)
\nonumber\\
&&-t_a
\left[
v_+^{(0)*}(k_x,k_y^0+\pi\Phi m)u_+^{(0)}(k_x,k_y^0+\pi\Phi(m-1))
\right]
\alpha_{m-1}(k_x,k_y^0)
\nonumber\\ 
&&-t_a
\left[
v_+^{(0)*}(k_x,k_y^0+\pi\Phi m)u_-^{(0)}(k_x,k_y^0+\pi\Phi(m-1)) 
\right]
\beta_{m-1}(k_x,k_y^0),
\label{eq:alpha}
\end{eqnarray}
where $\alpha_m(k_x,k_y^0)$ and $\beta_m(k_x,k_y^0)$ are defined as
\begin{eqnarray}
\alpha_m(k_x,k_y^0)=\alpha(k_x,k_y^0+\pi\Phi m),
\quad
\beta_m(k_x,k_y^0)=\beta(k_x,k_y^0+\pi\Phi m),
\end{eqnarray}
and $k_y^0$ satisfies $0\le k_y^0<2\pi/q$.
Here we have used the normalization conditions of 
the functions $(u_+^{(0)}({\bm k}),v_+^{(0)}({\bm k}))$ and
$(u_-^{(0)}({\bm k}),v_-^{(0)}({\bm k}))$.
In a similar manner, we also have
\begin{eqnarray}
&&\left(E+E^{(0)}(k_y^0+\pi\Phi m)\right)\beta_m(k_x,k_y^0)=
\nonumber\\
&&-t_a
\left[
u_-^{(0)*}(k_x,k_y^0+\pi\Phi m)v_+^{(0)}(k_x,k_y^0+\pi\Phi(m+1)) 
\right]
\alpha_{m+1}(k_x,k_y^0)
\nonumber\\
&&-t_a
\left[
u_-^{(0)*}(k_x,k_y^0+\pi\Phi m)v_-^{(0)}(k_x,k_y^0+\pi\Phi(m+1)) 
\right]
\beta_{m+1}(k_x,k_y^0)
\nonumber\\
&&-t_a
\left[
v_-^{(0)*}(k_x,k_y^0+\pi\Phi m)u_+^{(0)}(k_x,k_y^0+\pi\Phi(m-1))
\right]
\alpha_{m-1}(k_x,k_y^0)
\nonumber\\ 
&&-t_a
\left[
v_-^{(0)*}(k_x,k_y^0+\pi\Phi m)u_-^{(0)}(k_x,k_y^0+\pi\Phi(m-1)) 
\right]
\beta_{m-1}(k_x,k_y^0).
\label{eq:beta}
\end{eqnarray}
When $t_a=0$, these equations reduce to
\begin{eqnarray}
(E-E^{(0)}(k_y^0+\pi \Phi m))\alpha_m(k_x,k_y^0)=0, 
\nonumber\\
(E+E^{(0)}(k_y^0+\pi \Phi m))\beta_m(k_x,k_y^0)=0, 
\end{eqnarray}
and the unperturbed solutions are given by
\begin{eqnarray}
E=E^{(0)}(k_y^0+\pi \Phi m),
\quad
\alpha_m(k_x,k_y^0)=1, 
\quad
\beta_m(k_x,k_y^0)=0,
\end{eqnarray}
and
\begin{eqnarray}
E=-E^{(0)}(k_y^0+\pi \Phi m),
\quad
\alpha_m(k_x,k_y^0)=0, 
\quad
\beta_m(k_x,k_y^0)=1,
\end{eqnarray}
which reproduce (\ref{eq:u+0v+0}) and (\ref{eq:u-0v-0}), respectively.
In the new basis, $\alpha({\bm k})$ and $\beta({\bm k})$
describe the upper $(E>0)$ and lower $(E<0)$ bands, respectively.
The energy spectrum for $t_a=0$ is shown in Fig.\ref{fig:t=0}. 

In (\ref{eq:alpha}) and (\ref{eq:beta}), the terms proportional to $t_a$ give
couplings between the momenta $k_y$ and $k_y\pm \pi \Phi$.
So, if we put $\Phi=p/q$ with co-prime integers $p$ and $q$ $(q>0)$,
then in general gaps open when $E^{(0)}(k_y)=E^{(0)}(k_y')$ with
\begin{eqnarray}
k_y=k_y'+\pi \left(\frac{p}{q}\right)t,
\quad
k_y=-k_y'+\pi s. 
\label{eq:kyky'}
\end{eqnarray}
Here $s$ is an integer, and $k_y$ and $k_y'$ couple by $|t|$th-order
perturbation.  
The size of the gap is an order of $t_a^{|t|}$. 
If $s$ is chosen appropriately, $k_y$ is put between $0$ and $\pi$ and
$k_y$ for the $r$-th gap from the bottom
can be chosen as
\begin{eqnarray}
k_y=\left(\frac{r}{2q}\right)\pi, 
\label{eq:ky}
\end{eqnarray}
where $r$ is an integer with $1\le r \le 2q-1$.
(Gaps with $1\le r \le q-1$ open in the lower band $(E<0)$, and
those with $q+1\le r \le 2q-1$ open in the upper band $(E>0)$.
A gap at $E=0$ corresponds to $r=q$, but it closes when $\xi\rightarrow
0$ for a weak $t_a$.)
Eliminating $k_y$ and $k_y'$ from (\ref{eq:kyky'}) and (\ref{eq:ky}), we
have the Diophantine equation
\begin{eqnarray}
r=qs_r+pt_r,
\label{eq:D}
\end{eqnarray}
where we have written the subscript $r$ since $t$ and $s$ depend on
$r$ implicitly as a solution of (\ref{eq:D}).

\begin{figure}[h]
\begin{center}
\includegraphics[width=6cm]{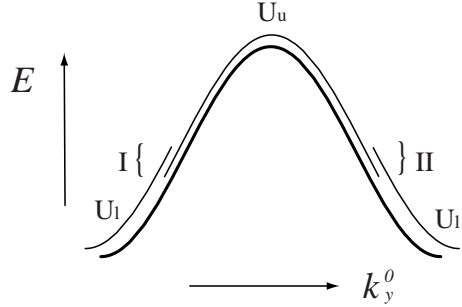}
\caption{An upper edge ${\rm U_u}$ and a lower one ${\rm U_l}$ of a
 band. They overlaps in the regions ${\rm I}$ and ${\rm II}$. }
\label{fig:patch}
\end{center}
\end{figure}

We can also derive the formula (\ref{eq:TKNN}) in the perturbation
theory. 
For brevity, we derive it only for gaps in the upper band ($E>0$).
The same formula can be obtained for gaps in the lower band ($E<0$) in a similar manner.
The wave function in the upper band ($E>0$) is simply 
given by
\begin{eqnarray}
\left[
\begin{array}{c}
u(k_x,k_y^0) \\
v(k_x,k_y^0)
\end{array}
\right] 
=\alpha_m(k_x,k_y^0)
\left[
\begin{array}{c}
u_+^{(0)}(k_x,k_y^0+\pi \phi m) \\
u_-^{(0)}(k_x,k_y^0+\pi \phi m)
\end{array}
\right]
+\alpha_{m'}(k_x,k_y^0)
\left[
\begin{array}{c}
u_+^{(0)}(k_x,k_y^0+\pi \phi m') \\
u_-^{(0)}(k_x,k_y^0+\pi \phi m')
\end{array}
\right],
\end{eqnarray}
where $m=m'+t$ and we take $0\le k_y^0<\pi/q $. The gaps are at $k_y^0=0$
and $\pi/2q$.
Write $\alpha_m=a$ and $\alpha_{m'}=b$, then an effective Schr\"{o}dinger
equation is given by
\begin{eqnarray}
\left[
\begin{array}{cc}
\epsilon & \eta e^{ik_x t}\\
\eta e^{-ik_x t}& -\epsilon
\end{array}
\right] 
\left[
\begin{array}{c}
a \\
b
\end{array}
\right]
=E
\left[
\begin{array}{c}
a \\
b
\end{array}
\right],
\label{eq:eSe}
\end{eqnarray}
where $\epsilon$ and $E$ are measured from the middle of the gap, namely
\begin{eqnarray}
E=E^{(0)}(k_y^0+\pi \phi m), 
\end{eqnarray}
and $k_y^0=0$ for even $r$ or $\pi/2q$ for odd $r$.
Here the $k_x$-dependence of the off diagonal elements of the
matrix in (\ref{eq:eSe}) is determined by (\ref{eq:u+0v+0}),
(\ref{eq:u-0v-0}) and (\ref{eq:alpha}), and the parameter $\eta$ is a
real number and of the order of $t_a^{|t|}$. 
The solutions of (\ref{eq:eSe}) are 
\begin{eqnarray}
&&E_+=\sqrt{\epsilon^2+\eta^2},
\nonumber\\
&&
\left[
\begin{array}{c}
a_+ \\
b_+
\end{array}
\right]
=\frac{1}{\sqrt{2E_+(E_+-\epsilon)}}
\left[
\begin{array}{c}
\eta \\
(E_+-\epsilon)e^{-ik_x t}
\end{array}
\right],
\label{eq:e+}
\end{eqnarray}
and 
\begin{eqnarray}
&&E_-=-\sqrt{\epsilon^2+\eta^2},
\nonumber\\
&&
\left[
\begin{array}{c}
a_- \\
b_-
\end{array}
\right]
=\frac{1}{\sqrt{2E_-(E_--\epsilon)}}
\left[
\begin{array}{c}
\eta \\
(E_--\epsilon)e^{-ik_x t}
\end{array}
\right].
\label{eq:e-}
\end{eqnarray}
When $\epsilon /|\eta| \rightarrow \infty$, these solutions behave as
\begin{eqnarray}
\left[
\begin{array}{c}
a_+ \\
b_+
\end{array}
\right]
\rightarrow
\left[
\begin{array}{c}
{\rm sgn}\eta \\
0
\end{array}
\right],
\quad
\left[
\begin{array}{c}
a_- \\
b_-
\end{array}
\right]
\rightarrow
\left[
\begin{array}{c}
0 \\
-e^{-ik_x t}
\end{array}
\right],
\label{eq:asymp1}
\end{eqnarray}
and when $\epsilon/|\eta| \rightarrow -\infty$,
\begin{eqnarray}
\left[
\begin{array}{c}
a_+ \\
b_+
\end{array}
\right]
\rightarrow
\left[
\begin{array}{c}
0 \\
e^{-i k_x t}
\end{array}
\right],
\quad
\left[
\begin{array}{c}
a_- \\
b_-
\end{array}
\right]
\rightarrow
\left[
\begin{array}{c}
{\rm sgn}\eta \\
0
\end{array}
\right].
\label{eq:asymp2}
\end{eqnarray}
For the upper edges of $r$-th band, we take the solution (\ref{eq:e-})
with $t=t_r$.
The asymptotic behaviors (\ref{eq:asymp1}) and (\ref{eq:asymp2}) imply that as
$k_y^0$ passes $0$ (for even $r$) or $\pi/2q$ (for odd $r$), the
wave function changes from
\begin{eqnarray}
\left[
\begin{array}{c}
u(k_x,k_y^0) \\
v(k_x,k_y^0)
\end{array}
\right]_- 
=
{\rm sgn}\eta
\left[
\begin{array}{c}
u_+^{(0)}(k_x,k_y^0+\pi \phi m) \\
u_+^{(0)}(k_x,k_y^0+\pi \phi m)
\end{array}
\right]
\label{eq:up2}
\end{eqnarray}
to
\begin{eqnarray}
\left[
\begin{array}{c}
u(k_x,k_y^0) \\
v(k_x,k_y^0)
\end{array}
\right]_- 
=
-e^{-ik_x t_r}
\left[
\begin{array}{c}
u_+^{(0)}(k_x,k_y^0+\pi \phi m') \\
u_+^{(0)}(k_x,k_y^0+\pi \phi  m' )
\end{array}
\right]
\label{eq:up1}.
\end{eqnarray}
And for the lower edge of $r$-th band, we take the solution (\ref{eq:e+})
with $t=t_{r-1}$, then
we have asymptotic behaviors of the wave function in a similar manner. 
On the overlap in the center of the band, these wave functions are
related to each other by transition functions.
See Fig.\ref{fig:patch}.
On the overlap ${\rm I}$ in Fig.\ref{fig:patch}, the wave function
(\ref{eq:up2}) is identified with
\begin{eqnarray}
\left[
\begin{array}{c}
u(k_x,k_y^0) \\
v(k_x,k_y^0)
\end{array}
\right]_+ 
=
{\rm sgn}\eta
\left[
\begin{array}{c}
u_+^{(0)}(k_x,k_y^0+\pi \phi m) \\
u_+^{(0)}(k_x,k_y^0+\pi \phi m)
\end{array}
\right],
\end{eqnarray}
by a trivial transition function, $e^{i\theta}=1$,
and on the overlap ${\rm II}$, (\ref{eq:up1}) is identified with
\begin{eqnarray}
\left[
\begin{array}{c}
u(k_x,k_y^0) \\
v(k_x,k_y^0)
\end{array}
\right]_+
=
e^{-ik_x t_{r-1}}
\left[
\begin{array}{c}
u_+^{(0)}(k_x,k_y^0+\pi \phi m') \\
u_+^{(0)}(k_x,k_y^0+\pi \phi  m' )
\end{array}
\right],
\end{eqnarray}
by the transition function
\begin{eqnarray}
e^{i\theta(k_x)}=-e^{ik_x (t_{r}-t_{r-1})}. 
\label{eq:tf}
\end{eqnarray}

Now let us calculate the contribution of the $r$-th band to the Hall
conductance,
\begin{eqnarray}
\sigma^{(r)}_{xy}&=&
-\frac{e^2}{h}\frac{1}{2\pi }\int \int dk_x dk_y^0
(\nabla_{\bm k}\times {\bm A}({\bm k}))_z.
\label{eq:chern}
\end{eqnarray}
Here the integration of (\ref{eq:chern}) is performed on the first magnetic
Brillouin zone, and ${\bm A}({\bm k})$ is defined as
\begin{eqnarray}
{\bm A}({\bm k})=-i[u^{*}({\bm k})\nabla_{\bm k}u({\bm k})
+v^{*}({\bm k})\nabla_{\bm k}v({\bm k})], 
\label{eq:gaugefield}
\end{eqnarray}
where $(u({\bm k}), v({\bm k}))$ is the wave function for the $r$-th band
constructed above. 
For the present purpose,
it is convenient to consider the rectangle region given by $-\pi\le k_x
<\pi$ and $0\le k_y^0 <2\pi/q$, which is equivalent to a pair of the
first magnetic Brillouin zone. 
Performing the integral (\ref{eq:chern}) in this region and dividing it
by 2, we have
\begin{eqnarray}
\sigma^{(r)}_{xy}=
-\frac{e^2}{h}\frac{1}{4\pi }\int_{-\pi}^{\pi}  d k_x \int_0^{2\pi /q}dk_y^0 
(\nabla_{\bm k}\times {\bm A}({\bm k}))_z.
\label{eq:chern2}
\end{eqnarray}
From the Stokes theorem, one can show that only the transition function
(\ref{eq:tf}) in the center of the band contributes to this integral.
Thus we obtain
\begin{eqnarray}
\sigma^{(r)}_{xy}=-\frac{e^2}{h}\frac{1}{4\pi }\int_{-\pi}^{\pi}d k_x \partial_{x}
\theta(k_x) \times 2  =-\frac{e^2}{h}(t_r-t_{r-1}). 
\end{eqnarray}
(The factor 2 arises because we have the transition function
(\ref{eq:tf}) twice in the integral region of (\ref{eq:chern2}).)
When the Fermi surface is in the $r$-th gap from the bottom, we have 
the formula (\ref{eq:TKNN}),
\begin{eqnarray}
\sigma_{xy}=\sum_{r'=1}^{r}\sigma_{xy}^{(r')}=-\frac{e^2}{h}t_r.
\end{eqnarray}

\subsection{Hall conductance and the Diophantine equation}
\label{subsec:HCDE}

Let us now determine the Hall conductance. 
As we showed above, the Hall conductance in units of $-e^2/h$, namely $t_r$,
satisfies the Diophantine equation. By combining with the perturbation
theory, the Diophantine equation enables us to determine 
the Hall conductance algebraically.

Before giving a rule to determine the Hall conductance, 
we briefly summarize properties of the Diophantine equation
which we shall use later.
First, the Diophantine equation has an infinite number of
solutions: If we have a solution
$(s_r,t_r)$ for a given $r$, then $(s_r-lp, t_r+lq)$
for any integer $l$ is also a solution.  
Second, the minimal value of $|t_r|$ among the solutions satisfies
$|t_r|\le q/2$.
We denote a solution with the minimal $|t_r|$ by $(s_r^0,t_r^0)$.
Third, except when $q$ is an even integer and $|t_r|=q/2$, $(s_r^{0},
t_r^0)$ is uniquely determined. 
When $|t_r|=q/2$, we have two solutions: $t_r^0=\pm q/2$.

As we denoted in Sec.\ref{subsec:WTP}, 
$|t_r|$ is an order of the perturbation.
So one naively expect that the size of the $r$th gap from the bottom and
its Hall conductance are
given by its minimal value $t_r^0$. 
But this is not always correct.
The reason why is that 
if an intermediate state in the $|t_r^0|$th order perturbation has the
momentum ${\bm k}$ with $\Delta^{(0)}({\bm k})=0$, (namely ${\bm k}$ with
$k_y=\pi/2$ $(\mbox{mod $\pi$})$, then
the transition amplitude becomes zero. 
For example, let us consider a gap at $k_y=(r/2q)\pi$ with
$r=q+4p$. (We assume that $q\gg p$.) 
\begin{figure}[h]
\begin{center}
\includegraphics[width=8cm]{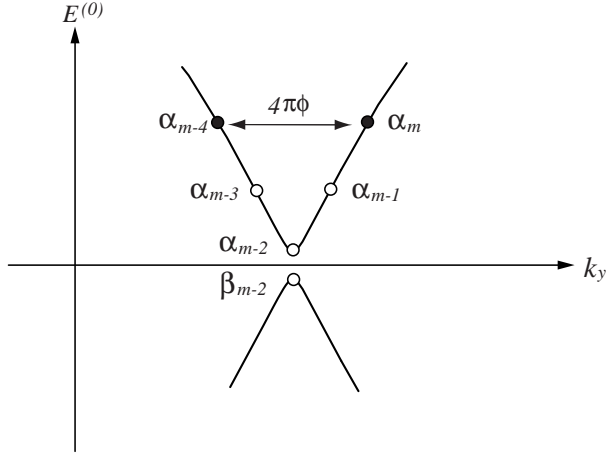}
\caption{Intermediate states for a gap at $k_y=(r/2q)\pi$ with $r=q+4p$
 in fourth order perturbation. The states $\alpha_m$ and
 $\alpha_{m'=m-4}$ represented by solid circles are mixed by the
 perturbation. Open circles represent intermediate states. }
\label{fig:even}
\end{center}
\end{figure}
In this case, the minimal value of $|t_r|$ is given by $t_r=4$, 
so a naive expectation is that the gap opens in fourth order
perturbation.
In Fig.\ref{fig:even},
we illustrate intermediate states.
As is seen there, we have to pass through either
$\alpha_{m-2}$ or $\beta_{m-2}$ as an intermediate state, and both of
them have $k_y=\pi/2$.
From (\ref{eq:alpha}) and (\ref{eq:beta}), the transition amplitude to
the intermediate state $\alpha_{m-2}$ is proportional to 
\begin{eqnarray}
v_+^{(0)}(k_x,k_y^0+\pi\Phi(m-2))
\propto \Delta^{(0)*}(k_x,\frac{\pi}{2}), 
\end{eqnarray} 
and that from $\beta_{m-2}$ is proportional to
\begin{eqnarray}
u_-^{(0)*}(k_x,k_y^0+\pi\Phi(m-2))
\propto \Delta^{(0)*}(k_x,\frac{\pi}{2}),  
\end{eqnarray}
and both of them are zero since $\Delta^{(0)}({\bm k})=0$ at $k_y=\pi/2$.
So the matrix element between $\alpha_m$ and $\alpha_{m-4}$ vanishes. 
As a result, no gap opens in the fourth order perturbation. 
The gap opens in a higher order perturbation.

Now let us derive a condition that the transition amplitude in
$|t_r^0|$th order perturbation vanishes. 
Without losing generality we can assume that $t_r^0\neq 0$ 
since $t_r^0$ becomes zero only for a gap at $E=0$, which
closes when $\xi\rightarrow 0$.
We also assume that $|t_r^0|\neq q/2$.
The case of $|t_r^0|=q/2$ is discussed later.
Consider the mixing between $k_y$ and $k_y'$ with
\begin{eqnarray}
k_y=\left(\frac{r}{2q}\right)\pi, 
\quad 
k_y=k_y'+\pi\left(\frac{p}{q}\right)t_r^0,
\quad
k_y=-k_y'+\pi s_r^0. 
\end{eqnarray}
There are $|t_r^0|-1$ intermediate states,
and since $k_x$ and $k_x'$ are given by
\begin{eqnarray}
k_y=\frac{\pi}{2}s_r^{0}
 +\frac{\pi}{2}\left(\frac{p}{q}\right)t_r^{0},
\quad
k_y'=\frac{\pi}{2}s_r^{0}
 -\frac{\pi}{2}\left(\frac{p}{q}\right)t_r^{0},
\end{eqnarray}
they have momenta
\begin{eqnarray}
k_y''=\frac{\pi}{2}s_r^0
+ \frac{\pi}{2}\left(\frac{p}{q}\right)(t_r^0-2l), 
\label{eq:momentum_inter}
\end{eqnarray}
where
\begin{eqnarray}
l=\left\{
\begin{array}{ll}
1,2,\cdots, t_r^0-1, & \mbox{for $t_r^0>0$},\\
-1,-2,\cdots, t_r^0+1, &\mbox{for $t_r^0<0$}.
\end{array}
\right.
\end{eqnarray}
If $k_y''$ in (\ref{eq:momentum_inter}) becomes $\pi /2$ $(\mbox{mod
$\pi$})$, then the transition amplitude in the $|t_r^0|$th order 
perturbation vanishes.
In order for $k_y''$ to be $\pi/2$ $(\mbox{mod $\pi$})$, 
\begin{eqnarray}
\left(\frac{p}{q}\right)(t_r^0-2l) 
\end{eqnarray}
must be an integer at least. 
And to satisfy this $t_r^0-2 l$ is needed to be a multiple of $q$ 
since $p$ and $q$ are co-prime integers.
But because $t_r^0$ satisfies 
\begin{eqnarray}
|t_r^0-2l|\le |t_r|\le \frac{q}{2},
\end{eqnarray}
this is possible only
when $t_r^0-2 l=0$.
Thus $t_r^0$ must be even.
Furthermore, 
$s_r^0$ must be odd in order for $k_y''$ to be $\pi/2$ $(\mbox{mod $\pi$})$
since $k_y''=(\pi/2)s_r^0$ when $t_r^0=2 l$.
Therefore only when $t_r^0$ is an even integer and $s_r^0$ an odd one, 
the transition amplitude vanishes.
Otherwise, we have a nonzero transition amplitude in the
$|t_r^0|$th order perturbation in general.

When $t_r^0$ is an even integer and $s_r^0$ an odd one,
a nonzero transition amplitude is obtained in the next leading order.
The next minimum value of $|t_r|$ is given by
\begin{eqnarray}
t_r=\left\{
\begin{array}{ll}
t_r^0-q, & \mbox{for $t_r^0>0$},\\
t_r^0+q, & \mbox{for $t_r^0<0$}.
\end{array}
\right. 
\end{eqnarray}
The intermediate states in this order have momenta 
\begin{eqnarray}
k_y''&=&
\frac{\pi}{2}s_r+\frac{\pi}{2}\left(\frac{p}{q}\right)(t_r-2l)
\nonumber\\
&=&
\frac{\pi}{2}s_r^0
+\frac{\pi}{2}\left(\frac{p}{q}\right)(t_r^0-2l) 
\end{eqnarray}
with $s_r=s_r^0+p$ $(s_r=s_r^0-p)$ for $t_r^0>0$ $(t_r^0<0)$ and  
\begin{eqnarray}
l=
\left\{
\begin{array}{ll}
-1,-2,\cdots, t_r^0-q+1, & \mbox{for $t_r^0>0$},\\
1,2,\cdots, t_r^0+q-1, & \mbox{for $t_r^0<0$}. 
\end{array}
\right. 
\end{eqnarray}
This $k_y''$ can not be $\pi/2$ $(\mbox{mod $\pi$})$:
In order for $k_y''$ to be $\pi/2$ $(\mbox{mod $\pi$})$,  
$t_r^0/2-l$ must be a multiple of $q$, but
this can not be met from the restriction $0<|t_r^0/2-l|<q$.

Now we summarize a procedure to determine the Hall conductance for the
$r$th gap from below under a magnetic field $\Phi=p/q$.
\begin{enumerate}
 \item First, find a solution of the Diophantine equation, 
$(s_r^0, t_r^0)$, which satisfies $|t_r^0|\le q/2$.
\item Then if $t_r^0$ is an even integer and $s_r^0$ an odd one,
the Hall conductance is given by
\begin{eqnarray}
\sigma_{xy}=\left\{
\begin{array}{ll}
-\frac{e^2}{h}(t_r^0-q), &\mbox{for $t_r^0>0$}, \\
-\frac{e^2}{h}(t_r^0+q), &\mbox{for $t_r^0<0$},
\end{array}
\right.
\label{eq:rule1}
\end{eqnarray} 
The size of the gap is an order $t_a^{q-|t_r^0|}$.
\item For the other cases, 
the Hall conductance is given by
\begin{eqnarray}
\sigma_{xy}=-\frac{e^2}{h}t_r^0,
\label{eq:rule2}
\end{eqnarray}
and the size of the gap is an order of $t_a^{|t_r^0|}$.
\end{enumerate}

Finally, let us discuss the case of $|t_r^0|=q/2$. 
This is possible only when $q$ is even and only for gaps with
$r=q/2$ or $r=3q/2$. 
Although we have two solutions of the Diophantine equation,
$t_r^0=\pm q/2$, the Hall conductance is determined uniquely if
$q/2$ is an even integer.
Since $p$ is odd when $q$ is even,
it can be shown that one of these solutions has an odd
$s_r^0$ and the other has an even $s_r^0$.
So from the argument above, it is found that if $q/2$ is an even number, 
the solution with an odd $s_r^0$ has vanishing transition amplitude in
the $(q/2)$-th perturbation.
Thus the Hall conductance is given by the solution $t_r^0$ with an
even $s_r^0$.

When $q/2$ is an odd number, both of the solutions with $t_r^0=\pm q/2$
have non zero transition amplitudes. 
For $q=2$, the gap closes since the transition amplitudes of these
solutions are the same up to a phase and cancel each other at some $k_x$, 
but for the other case, (namely $q=2(2n+1)$ with integers $n$), a gap
opens and the Hall conductance is given by $t_r^0$ with a larger
amplitude,
which can not be determined algebraically.

\subsection{gap structure and unconventional Hall conductance in a weak
  magnetic field} 
\label{subsec:UHCWMF}

Let us now examine the Hall conductance for the weak $t_a$ case in a weak
magnetic field. 
Consider the $r(=qs_r+p t_r)$th gap from the bottom.
Since $r$ satisfies $0<r<2q$,
$t_r$ is restricted to
\begin{eqnarray}
-\left(\frac{q}{p}\right)s_r < t_r < (2-s_r)\left(\frac{q}{p}\right). 
\end{eqnarray}
So if $s_r\ge 3$ or $s_r\le -1$,  
$|t_r|$ is bounded from below by $(q/p)$. 
In these cases, the size of the gap $O(t_a^{|t_r|})$ becomes zero in a weak magnetic
field limit, $(p/q)\rightarrow 0$.
Therefore, visible gaps in a weak magnetic field are possible only for
the following three classes of $r$, (a) $r=p t_r$, $(t_r=1,2,\cdots, )$,
(b) $r=2q+p t_r$ $(t_r=-1,-2,\cdots,)$, and (c) $r=q+p t_r$ $(t_r=\pm
1,\pm 2,\cdots,)$.

In the case of (a), 
we have $(s_r^0,t_r^0)=(0,n)$ if we put $t_r=n$ with $0<n< q/2$.
Since $s_r^0$ is even, the Hall conductance for $r=pn$th gap from below
is given by
\begin{eqnarray}
\sigma_{xy}=-\frac{e^2}{h}n,
\quad (n=1,2,\cdots).
\end{eqnarray}
Thus the conventional quantization holds in this
case. These gaps are near the bottom of the spectrum.
In a similar manner, it can be shown that gaps in the class (b) also
show the conventional quantization for the Hall conductance.
Namely the Hall conductance for the gap with $r=2q-pn'$
($ 0<n'<q/2$) is given by
\begin{eqnarray}
\sigma_{xy}=\frac{e^2}{h}n',
\quad (n'=1,2,\cdots,).
\end{eqnarray}
These gaps are seen near the top of the spectrum.

For visible gaps in the class (c), we have the unconventional Hall
conductance. If we put $t_r=n''$ with $|n''|<q/2$, we have
an odd $s_r^0$. ($(s_r^0,t_r^0)=(1,n'')$.) 
Thus we have a visible gap 
only if $n''$ is an odd number, $n''=2m+1$ ($m$ is an integer). 
Otherwise, the lowest order perturbation is an order of
$q$, then the corresponding gap is invisible in a weak magnetic field limit.
For this class of visible gaps with $r=q+p(2m+1)$ and $|2m+1|<q/2$,
we have the unconventional quantization for the Hall conductance
\begin{eqnarray}
\sigma_{xy}=-\frac{e^2}{h}(2m+1),
\quad (m=0,\pm 1,\pm 2, \cdots,).
\end{eqnarray}
These gaps are located around $E=0$. 

\section{Strong $t_a$ phase}
\label{sec:STP}

\begin{figure}[ht]
\begin{center}
\includegraphics[width=6cm]{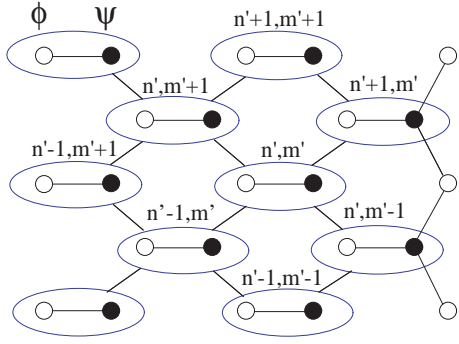}
\caption{The square lattice appearing in the strong $t_a$ limit of the
 honeycomb lattice in Fig.\ref{fig:Honeylattice}. Here $n'=(n+m)/2$ and
 $m'=(m-n)/2$.}
\label{fig:SquareLattice}
\end{center}
\end{figure}
Now let us consider the strong coupling limit, $t_a\gg 1$.
In this limit, the upper $(E>0)$ and the lower
($E<0$) parts of the spectrum separate from each other, and both of them  
are equivalent to that of the square lattice shown in
Fig.\ref{fig:SquareLattice}.
To see this, consider first the case of $\Phi=0$.
When $\Phi=0$, (\ref{eq:ftbm}) becomes
\begin{eqnarray}
\left(h({\bm k})-t_a\sigma_x \right)
\left[
\begin{array}{c}
u({\bm k}) \\
v({\bm k})
\end{array}
\right]
=E
\left[
\begin{array}{c}
u({\bm k}) \\
v({\bm k})
\end{array}
\right]. 
\label{eq:tbmphi=0}
\end{eqnarray}
In the leading order of the strong coupling limit, this equation reduces to
\begin{eqnarray}
-t_a\sigma_x 
\left[
\begin{array}{c}
u({\bm k}) \\
v({\bm k})
\end{array}
\right] 
= E \left[
\begin{array}{c}
u({\bm k}) \\
v({\bm k})
\end{array}
\right],
\end{eqnarray}
thus we have two flat bands
\begin{eqnarray}
\frac{1}{\sqrt{2}}
\left[
\begin{array}{c}
 1\\
-1
\end{array}
\right], 
\quad
E=t_a,
\end{eqnarray}
and 
\begin{eqnarray}
\frac{1}{\sqrt{2}}
\left[
\begin{array}{c}
 1\\
1
\end{array}
\right],
\quad 
E=-t_a.
\end{eqnarray}
In order to discuss the next leading order, it is convenient to take
the following basis,
\begin{eqnarray}
\left[
\begin{array}{c}
u({\bm k}) \\
v({\bm k})
\end{array}
\right] 
=a({\bm k})
\frac{1}{\sqrt{2}}
\left[
\begin{array}{c}
1 \\
-1
\end{array}
\right]
+
b({\bm k})
\frac{1}{\sqrt{2}}
\left[
\begin{array}{c}
1 \\
1
\end{array}
\right].
\end{eqnarray}  
In this basis, (\ref{eq:tbmphi=0}) is rewritten as
\begin{eqnarray}
\xi b({\bm k})
+\frac{\Delta^{(0)}({\bm k})-\Delta^{(0)*}({\bm k})}{2} b({\bm k})
-\frac{\Delta^{(0)}({\bm k})+\Delta^{(0)*}({\bm k})}{2} a({\bm k})
+t_a a({\bm k})=E a({\bm k}) 
\nonumber\\
\xi a({\bm k})
+\frac{\Delta^{(0)}({\bm k})+\Delta^{(0)*}({\bm k})}{2} b({\bm k})
-\frac{\Delta^{(0)}({\bm k})-\Delta^{(0)*}({\bm k})}{2} a({\bm k})
-t_a b({\bm k})=E b({\bm k}),
\end{eqnarray}
and we can neglect the mixing terms between the upper band
and lower one in the next leading order. Therefore,  we have
\begin{eqnarray}
-\frac{\Delta^{(0)}({\bm k})+\Delta^{(0)*}({\bm k})}{2} a({\bm k})
+t_a a({\bm k})=E a({\bm k}), 
\nonumber\\
\frac{\Delta^{(0)}({\bm k})+\Delta^{(0)}({\bm k})}{2} b({\bm k})
-t_a b({\bm k})=E b({\bm k}).
\end{eqnarray}
The eigen energies of these equations are $E=\pm (t_a + 2\cos k_x \cos k_y)$.
By changing variables as $k'_1=k_x+k_y$ and $k'_2=k_y-k_x$, they reduce
to the energies of a pair of the tight binding models on the square
lattice up to the constant shifts $\pm t_a$, 
\begin{eqnarray}
E=\pm (t_a+ \cos k'_1+\cos k'_2).
\end{eqnarray}

Now consider the case of $\Phi=p/q$.
For this purpose, we take a new gauge for the magnetic field and
consider the following equations,
\begin{eqnarray}
-e^{i\pi \Phi(n+m)}\phi_{n+1,m-1}-\phi_{n+1,m+1}-t_a \phi_{n,m}
+\xi \psi_{n,m} =E\psi_{n,m},
\nonumber\\
-e^{-i\pi\Phi(n+m)}\psi_{n-1,m+1}-\psi_{n-1,m-1}-t_a \psi_{n,m}
-\xi \phi_{n,m}=E\phi_{n,m}.
\end{eqnarray}
In terms of $a({\bm k})$ and $b({\bm k})$, these equations are rewritten
as  
\begin{eqnarray}
&&-\frac{e^{-ik'_2}}{2}
\left(b(k'_1-2\pi \Phi,k'_2)-a(k'_1-2\pi \Phi,k'_2)\right)
+\frac{e^{ik'_2}}{2}
\left(b(k'_1+2\pi\Phi,k'_2)+a(k'_1+2\pi\Phi,k'_2)\right)
\nonumber\\
&&+\cos k'_1 a({\bm k'})-i\sin k'_1 b({\bm k'})+t_a a({\bm k'})+\xi b({\bm k'})
=E a({\bm k'}),
\nonumber\\
&&-\frac{e^{-ik'_2}}{2}
\left(b(k'_1-2\pi \Phi,k'_2)-a(k'_1-2\pi \Phi,k'_2)\right)
-\frac{e^{ik'_2}}{2}\left(b(k'_1+2\pi\Phi,k'_2)+a(k'_1+2\pi\Phi,k'_2)\right)
\nonumber\\
&&-\cos k'_1 b({\bm k'})+i\sin k'_1 a({\bm k'})-t_a b({\bm k'})+\xi a({\bm k'})
=E b({\bm k'}),
\end{eqnarray}
where $(k_1',k_2')$ are momentum conjugate to the coordinate $(n', m')$,
which are given by $k_1'=k_x+k_y$ and $k_2'=k_y-k_x$.
In the leading order, we have again the two flat bands,
\begin{eqnarray}
t_a a({\bm k'})=E a({\bm k'}),
\quad
-t_a b({\bm k'})=E b({\bm k'}), 
\end{eqnarray}
and in the next leading order, the mixing terms between the upper
band ($E=t_a$) and lower one ($E=-t_a$) can be neglected as
\begin{eqnarray}
\frac{e^{ik'_2}}{2}a(k'_1+2\pi \Phi,k'_2)
+\frac{e^{-k'_2}}{2}a(k'_1-2\pi \Phi,k'_2)
+\cos k_1 a({\bm k'})=(E -t_a)a({\bm k'}),
\nonumber\\ 
-\frac{e^{ik'_2}
}{2}b(k'_1+2\pi \Phi,k'_2)
-\frac{e^{-ik'_2}}{2}b(k'_1-2\pi \Phi,k'_2)
-\cos k'_1 b({\bm k'})=(E+t_a) b({\bm k'}). 
\end{eqnarray}
These equations coincide with that of the tight-binding model on the
square lattice illustrated in Fig.\ref{fig:SquareLattice} 
except for the constant shift in the energies.

Using the results of the tight binding models on the square lattice in
\cite{TKNN82, Kohmoto88}, 
we can determine the Hall conductance for
$r$-th gap in the strong $t_a$ limit as 
\begin{eqnarray}
\sigma_{xy}=-\frac{e^2}{h}t_r^0,
\label{eq:Hc_strong}
\end{eqnarray}
where $t_r^0$ is a solution of the Diophantine equation with $|t_r^0|\le q/2$.
For $r=q/2$ and $r=3q/2$, we can not determine the Hall conductance
algebraically since we have two solutions $t_r^0=\pm q/2$. 
The corresponding gaps in the strong coupling limit
close, but for a strong but finite $t_a$ they open from a mixing between
the upper band and the lower one. The Hall conductance is given by
one of the two possible values of $t_a(=\pm q/2)$, but there is no
algebraic rule to determine it.

Let us comment briefly the Hall conductance for the strong $t_a$ case in a
weak magnetic field.
As was shown before, only three classes of gaps are visible in a weak
magnetic field: (a) $r=p t_r$, $(t_r=1,2,\cdots,)$, (b) $r=2q+p t_r$,
$(t_r= -1, -2,\cdots,)$, (c) $r=q+p t_r$, $(t_r=\pm 1,\pm 2,\cdots,)$.
For these classes of gaps, we have (a) $(s_r^0, t_r^0)=(0,n)$ $(0<n<q/2)$, (b)
$(s_r^0, t_r^0)=(2,-n')$ $(0<n'<q/2)$ and (c) $(s_r^0,t_r^0)=(1,n'')$
$(|n''|<q/2)$, respectively. 
In contrast to the weak $t_a$ case, the formula (\ref{eq:Hc_strong})
indicates that all of them show the conventionally quantized Hall conductance.

\section{phase transitions in intermediate $t_a$ region}
\label{sec:intermediate}
\subsection{topological quantum phase transitions }
\label{sec:TQPT}

In the above we have calculated the Hall conductance in two opposite
limits, $t_a\rightarrow 0$ and $t_a \rightarrow \infty$.
The algebraic rules to determine the Hall conductances in the both limits
were given and it was found that for gaps with even
$t_r^0$ and odd $s_r^0$ the Hall conductances take different values from
each other.
It implies that each gap in this class closes at some $t_a$ at least once
in order to change the corresponding topological number $t_r$.
In other words, we should have a quantum phase transition between
topological insulator phases at some $t_a$ if the
Fermi energy is located in a gap in this class.

In the following,
we will confirm the existence of the quantum phase transition numerically.
In addition, we will find that an unexpected topological quantum phase
transition occurs due to the van Hove singularity.

\subsection{numerical studies of quantum phase transitions}
\label{sec:NSQPT}

Here, we will study the whole region
of $t_a$ for a number of cases of $\Phi=p/q$ by using numerical calculations.  
We will find that except for a few gaps
either the strong $t_a$ limit or the weak $t_a$ one explains
the gap structures qualitatively and the Hall conductance
quantitatively: 
For a weak $t_a$ region, $0<t_a\le 1$, the gap structures 
are the same as those in the weak $t_a$ limit, and for a
strong $t_a$ region, $t_a>2$, they are the same as those in the strong
$t_a$ limit. 
In the intermediate region, $1<t_a \lesssim 2$, we have the topological
quantum phase transitions mentioned in the previous subsection and the gap
structure changes from those in the weak $t_a$ region to those in the
strong one. 
From the topological nature of the Hall conductance, these results
imply that except for a few gaps the Hall conductances in
the whole $t_a$ region are determined algebraically by using either
those in the weak $t_a$ region or those in the strong one.  

It will be shown that the exceptional gaps appear just beyond the van Hove singularity at
$t_a=1$ (namely the van Hove singularity in graphene) or on the lower
and the upper edges of the spectrum.
For some gaps in these regions, additional topological quantum phase
transitions, which are unexpected from the analysis in the two limits,
are observed.
It will be found that multiple topological quantum phase transitions
occur for these gaps. 

In the following we limit our calculation to $0\le \Phi\le 1/2$ without
losing generality: This is because the spectrum is invariant under
translation $\Phi\rightarrow\Phi+n$ with an integer $n$, and reflection
about half integer values of $\Phi$.
Furthermore a close inspection of the model also shows
that the spectrum is unchanged under reflection $E\leftrightarrow -E$
for a given $\Phi=p/q$.

\begin{figure}[ht]
\begin{center}
\includegraphics[width=7cm]{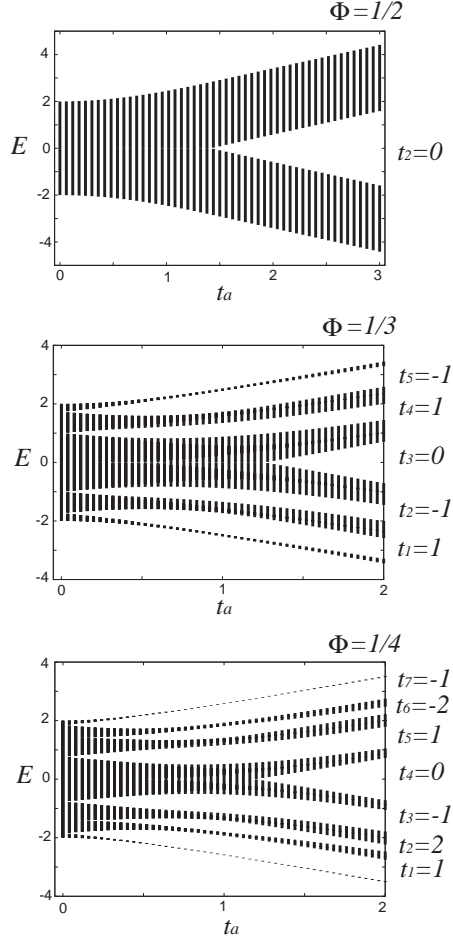}
\caption{The energy spectra v.s $t_a$ for $\Phi=1/2$, $1/3$, and
 $1/4$. None of gaps closes as we make $t_a$ stronger. 
The Hall conductance $\sigma_{xy}=-(e^2/h)t_r$ for $r$-th gap from the
 bottom, which is calculated algebraically by the method described in
 the text, is also shown.}
\label{fig:dat_p1q2+p1q3+p1q4}
\end{center}
\end{figure}

\begin{figure}[ht]
\begin{center}
\includegraphics[width=7cm]{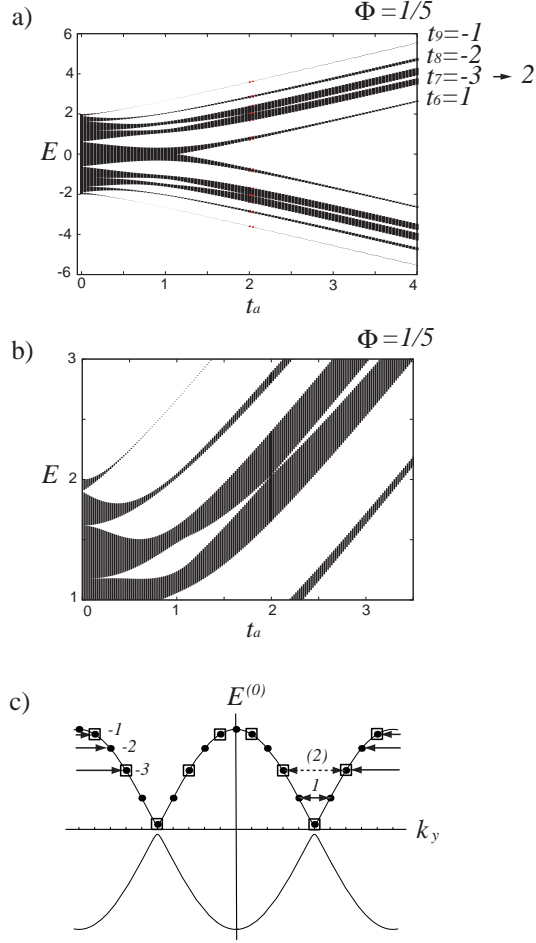}
\caption{a) The energy spectrum for $\Phi=1/5$. b) A closer look at the
 gap closing point in the seventh gap from the bottom. c) The unperturbed
 states and the mixing in the weak $t_a$ perturbation for $\Phi=1/5$.
The states with the same symbol are related to each other in the
 perturbation. The solid arrows indicate the mixing between  states.
 The dotted arrow indicates the mixing with vanishing amplitude,
 which has an intermediate state with the singular momentum $k_y=\pi/2$.
Integers denote $t_r$ for the mixing.}
\label{fig:dat_p1q5+t=0p1q5}
\end{center}
\end{figure}

\begin{figure}[ht]
\begin{center}
\includegraphics[width=7cm]{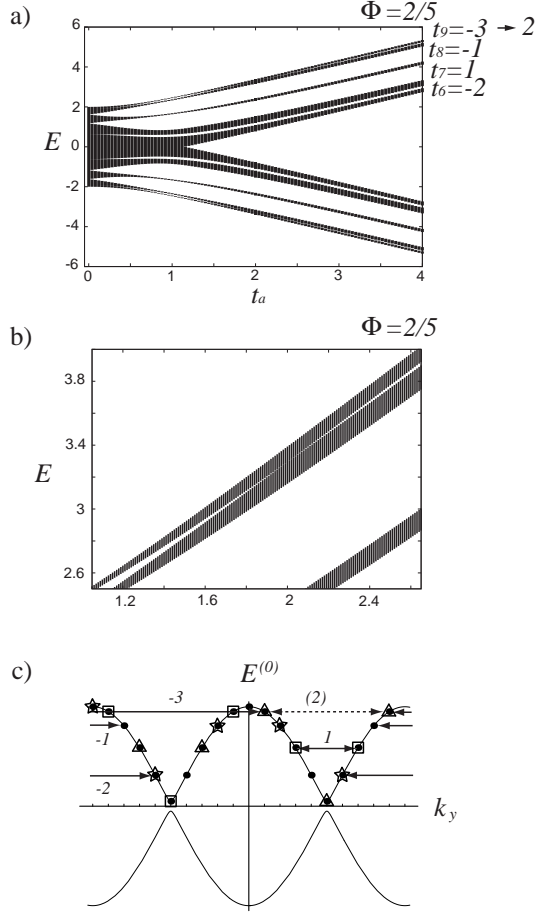}
\caption{a) The energy spectrum v.s $t_a$ for $\Phi=2/5$.
b) A closer look at the gap closing point in the ninth gap from the bottom.
c) The unperturbed states and the mixing in the weak $t_a$ perturbation
 for $\Phi=2/5$.}
\label{fig:dat_p2q5+t=0p2q5}
\end{center}
\end{figure}

\begin{figure}[h]
\begin{center}
\includegraphics[width=7cm]{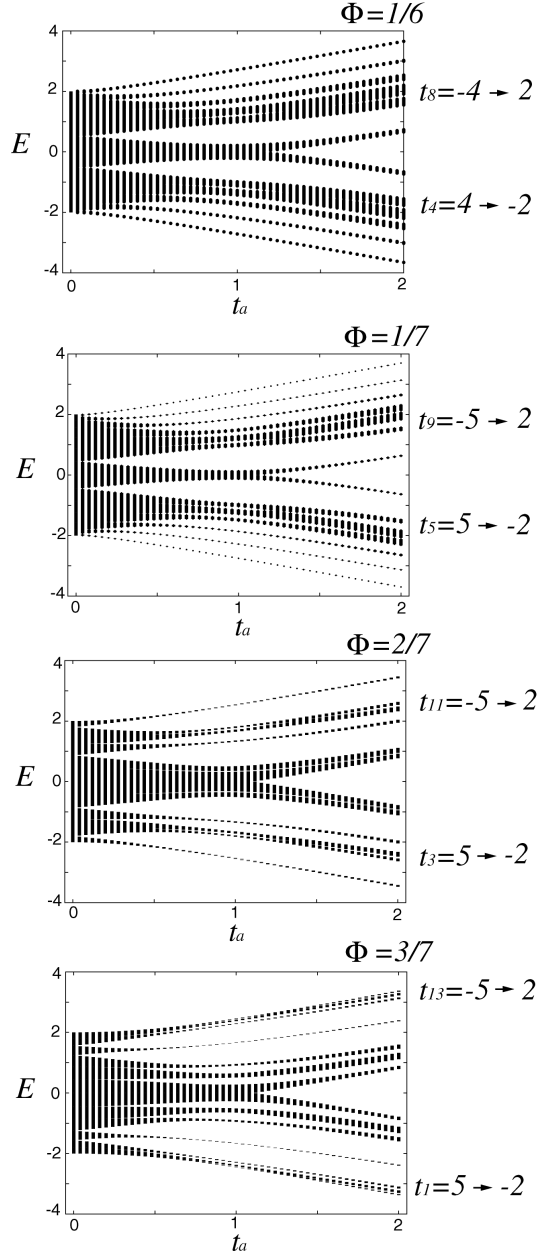}
\caption{The energy spectra v.s $t_a$ for $\Phi=1/6$, $1/7$, $2/7$ and $3/7$. 
The gap closing points appear whenever 
the Hall conductance ($\sigma_{xy}=-(e^2/h)t_r$) in the weak $t_a$
 region is different from that in the strong $t_a$ region.}
\label{fig:dat_p1q6+p1q7+p2q7+p3q7}
\end{center}
\end{figure}

\begin{figure}[ht]
\begin{center}
\includegraphics[width=7cm]{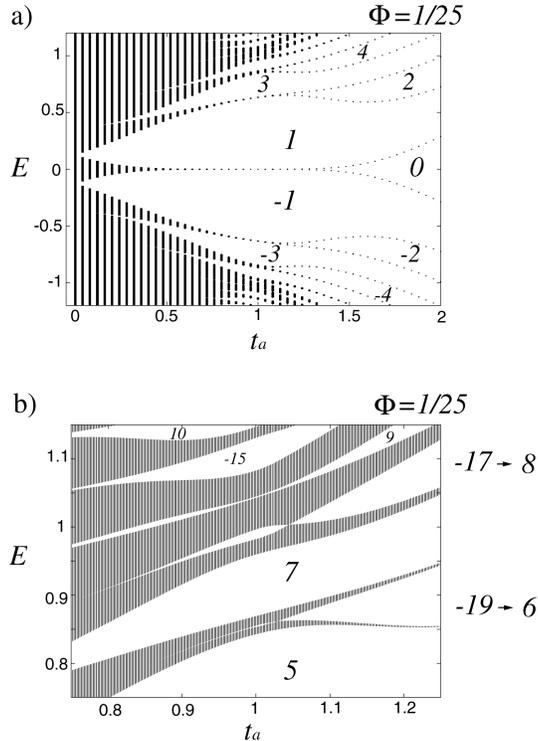}
\caption{a) The energy spectrum v.s $t_a$ for $\phi=1/25$. Integers in the gaps
indicates $t_r$ of the Hall conductance $\sigma_{xy}=-(e^2/h)t_r$. 
b) A closer look at gap closing points near $t_a=1$. It is found that gap
 closing points appear only when $t_a>1$. The gap with $t_r=-15$ closes
 at $t_a\sim 1.25$ (not shown).}
\label{fig:dat_p1q25}
\end{center}
\end{figure}

\begin{figure}[ht]
\begin{center}
\includegraphics[width=7cm]{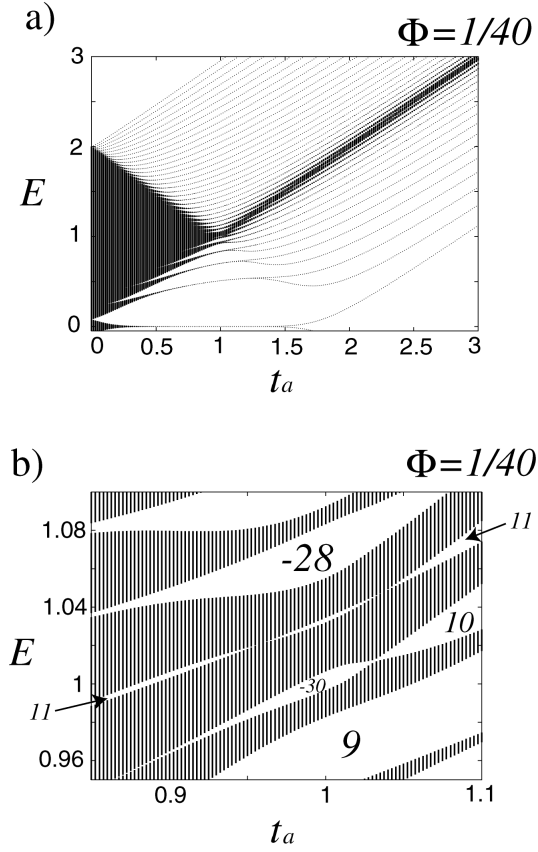}
\caption{a) The energy spectrum v.s $t_a$ for $\phi=1/40$.
b) A closer look at gap closing points near $t_a=1$. Integers in the
 gaps indicate $t_r$ of the Hall conductance $\sigma_{xy}=-(e^2/h)t_r$. 
 For the gap with $t_r^0=11$,  we have two gap closing points which are
 not required by the topological difference between the weak and strong
 $t_a$ limits.}
\label{fig:dat_p1q40}
\end{center}
\end{figure}

\begin{figure}[ht]
\begin{center}
\includegraphics[width=6cm]{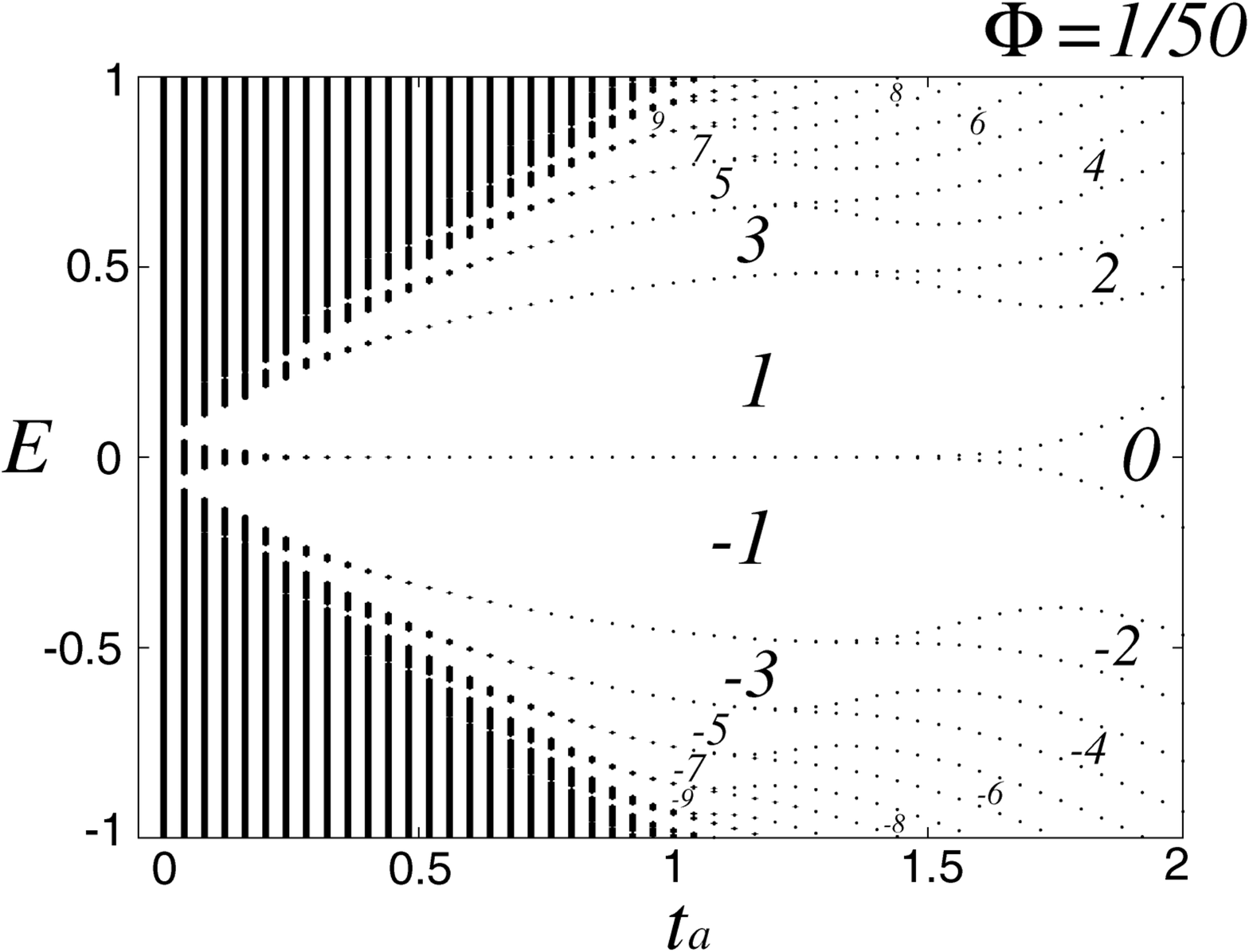}
\caption{
The energy spectrum v.s. $t_a$ for $\phi=1/50$. Integers indicate $t_r$
 of the Hall conductance $\sigma_{xy}=-(e^2/h)t_r$.}
\label{fig:dat_p1q50}
\end{center}
\end{figure}

We illustrate typical examples of energy spectra in
Figs.\ref{fig:dat_p1q2+p1q3+p1q4}-\ref{fig:dat_p1q50}.
For $\Phi=1/2, 1/3$ and
$1/4$, as is shown in Fig.\ref{fig:dat_p1q2+p1q3+p1q4}, none of the gaps
closes in a finite region of $t_a$, and a gap opens at $E=0$ when $t_a>1$. 
Since the Hall conductances for the unclosed gaps remain the same values
when $t_a$ changes,
they can be calculated algebraically by using the results of the weak
$t_a$ limit or those of the strong one.
The Hall conductances determined in
this manner are also shown in Fig.\ref{fig:dat_p1q2+p1q3+p1q4}.
For some gaps, both the weak and strong $t_a$ analyses can be used to
determine the Hall conductances and they are found to give
consistently the same values.

For $\Phi=1/5$ and $2/5$, there are gap closing points in the energy
spectra at $t_a\sim 2$.
See Figs.\ref{fig:dat_p1q5+t=0p1q5} and \ref{fig:dat_p2q5+t=0p2q5}.
It is found that for the gaps with the gap closing points, the Hall
conductances obtained in the weak $t_a$ analysis are different from
those in the strong one.
This means that the gap closing points are required by the topological
quantum phase transition described in Sec.\ref{sec:TQPT}.
In Figs.\ref{fig:dat_p1q5+t=0p1q5} and  \ref{fig:dat_p2q5+t=0p2q5} we
also illustrate intermediate states in the weak $t_a$ perturbation.
One can see that for the gaps with the gap closing points, there exists
the intermediate state with $k_y=\pm \pi/2$ $(\mbox{mod. $\pi$})$. 
For the other gaps we have no intermediate states with the these momenta.
Similar gap closing points are also found for $\Phi=1/6, 1/7, 2/7$ and $3/7$ in
Fig.\ref{fig:dat_p1q6+p1q7+p2q7+p3q7}.
These gap closing points are also required by the topological quantum
phase transition.
It is shown in Fig.\ref{fig:dat_p1q6+p1q7+p2q7+p3q7} how $t_r$ of the
Hall conductance $\sigma_{xy}=-(e^2/h)t_r$ changes from the weak $t_a$
region to the strong one.

When $q$ is large ($q\ge 7$), a new kind of gap closing
points, which are not required by the
topological difference between the weak and strong $t_a$ limits, appears
for some exceptional gaps.
For example, see Fig. \ref{fig:dat_p1q40}.
We have two gap closing points in the same gap where the quantum Hall
conductance in the weak $t_a$ limit is the same as that in the strong
$t_a$ limit. 
There is no a prior topological reason why the gap should close in
the intermediate $t_a$.
The possible origin of the additional gap closing points is accumulation
of gaps at these singularities.
For a large $q$,
many magnetic bands are accumulated both near the van Hove singularity at
$t_a=1$ and on the lower and upper edges of the spectrum, 
so the gaps near the singularities become narrow and are easy to close
as the accumulation grows.
Except near the singularities,
we did not observe the additional gap closing points

To illustrate the weak magnetic field case, $\Phi\ll 1$, 
we also show the energy spectra for $1/25$, $1/40$ and $1/50$ in
Figs.\ref{fig:dat_p1q25}, \ref{fig:dat_p1q40} and \ref{fig:dat_p1q50}.
The Hall conductances for visible gaps are also shown in these figures. 
It is found that in the weak $t_a$ region including the graphene case
($t_a=1$) the Hall conductances are unconventionally
quantized as $\sigma_{xy}=e^2(2n+1)/h$, $(n=0,\pm 1,\pm 2,\cdots)$ near
half-filling ($E= 0$), while in the strong $t_a$ region
they are conventionally quantized as $\sigma_{xy}=e^2 n/h$, $(n=0,\pm 1,
\pm 2,\cdots)$. 

\section{graphene case ($t_a=1$)}
\label{sec:graphene}
\subsection{Hall conductance}
\label{sec:HCG}

Let us now apply our results to graphene, where $t_a$ is given by $1$,
and determine its Hall conductances. 
As was shown in Sec.\ref{sec:NSQPT}, the numerical studies strongly
suggest that for most energy regions graphene ($t_a=1$) is
in the weak $t_a$ phase: Except the energy regions just beyond the van
Hove singularities at
$E=\pm 1$ (namely, the energy regions where $|E|$ is slightly greater
than $1$) or those on the lower and upper edges of the spectrum none of
gaps closes as we change $t_a$ from $0$ to $1$, and the structure of the
energy spectrum is qualitatively the same as that in the weak $t_a$ limit.
In particular, if we consider the $r$-th gap from the bottom, only the
gaps with the following three classes of $r$ are visible in a weak
magnetic field: (a) $r=pt_r$, $(t_r=1,2,\cdots,)$, (b) $r=2q+pt_r$, 
$(t_r=-1,-2,\cdots,)$ (c) $r=q+pt_r$, $(t_r=\pm 1, \pm 3, \pm 5,\cdots,)$.

These results indicates that even when $t_a=1$ the
perturbation theory works well at least qualitatively for gaps away from
the exceptional energy regions.
From the topological nature of the Hall conductance, it implies that
the Hall conductances in graphene for these gaps
are quantitatively the same as those at the weak $t_a$ limit. 
Namely they are determined algebraically 
by the rule (\ref{eq:rule1}) and (\ref{eq:rule2}).
In addition to the visible gaps in a weak magnetic field,
this rule also determines the Hall conductances for the subband gaps in
a strong magnetic field.

This result consistently explains the unconventional quantization of the Hall
conductance observed experimentally near half filling \cite{NMMFKZJSG05,ZTSK05}.
Among the visible gaps in a weak magnetic field, only 
those in the class (c) include the gaps near half
filling $(E=0)$.
They show the unconventional quantization of the Hall
conductance, $\sigma_{xy}=-e^2 t_r/h$ $(t_r=\pm 1,\pm 3, \cdots,)$.

We emphasize here that our result indicates the unconventional Hall
conductance persists up to the van Hove singularities, where the Dirac
fermion description is no longer valid. 
This is because none of gaps in this energy region closes
when we change $t_a$ from $0$ to $1$. 
They are well described by the weak $t_a$ analysis and
the Hall conductance for the visible gaps in a weak magnetic field shows
the unconventional quantization.
(Note that they belongs to the class (c) above.) 
This anomalous behavior of the Hall conductance on honeycomb lattice was
already reported in a numerical calculation \cite{HFA06}, but our study
here establishes this analytically.
Furthermore, at the same time, it reveals why the unconventional Hall
 conductance does not persist beyond the van Hove singularities:
For gaps just beyond the van Hove singularities, a gap closing point 
can appear when $t_a$ changes from $0$ to
$1$. So in general the weak $t_a$ analysis does not apply to the gaps in
this region and there is no reason why the unconventional Hall
conductances hold.  
The unconventional quantization of the Hall conductance up the the van Hove
singularities should be observed if the chemical potential can be varied
over a wide range in graphene.

\subsection{edge states}

To confirm the results in the previous section, we compare here
the bulk quantum Hall conductance obtained by the rule (\ref{eq:rule1}) and
(\ref{eq:rule2}) and the number of edge states obtained numerically for
finite systems with $t_a=1$.
From the bulk-edge correspondence in quantum Hall systems
\cite{Halperin82,Hatsugai93,Hatsugai93-2,QWZ06}, 
these quantum numbers should coincide with each other
for gaps with no gap closing point in the region $0<t_a\le 1$.
In other words, the results in the previous section suggest that they
should be the same except for some gaps just beyond the van Hove
singularities or those on the upper and lower edges in the spectrum. 

To calculate the number of edge states, we consider cylindrical
systems with bearded edges at both ends. 
For all possible $\Phi=p/q$ $(0<\Phi\le 1/2)$ with $1\le q
\le 11$, the energy spectra of these systems are obtained numerically
and the numbers of the edge states are evaluated.
We also examine the case of $\Phi=1/25$ as a weak
magnetic field case. 

\begin{figure}[ht]
\begin{center}
\includegraphics[width=8cm]{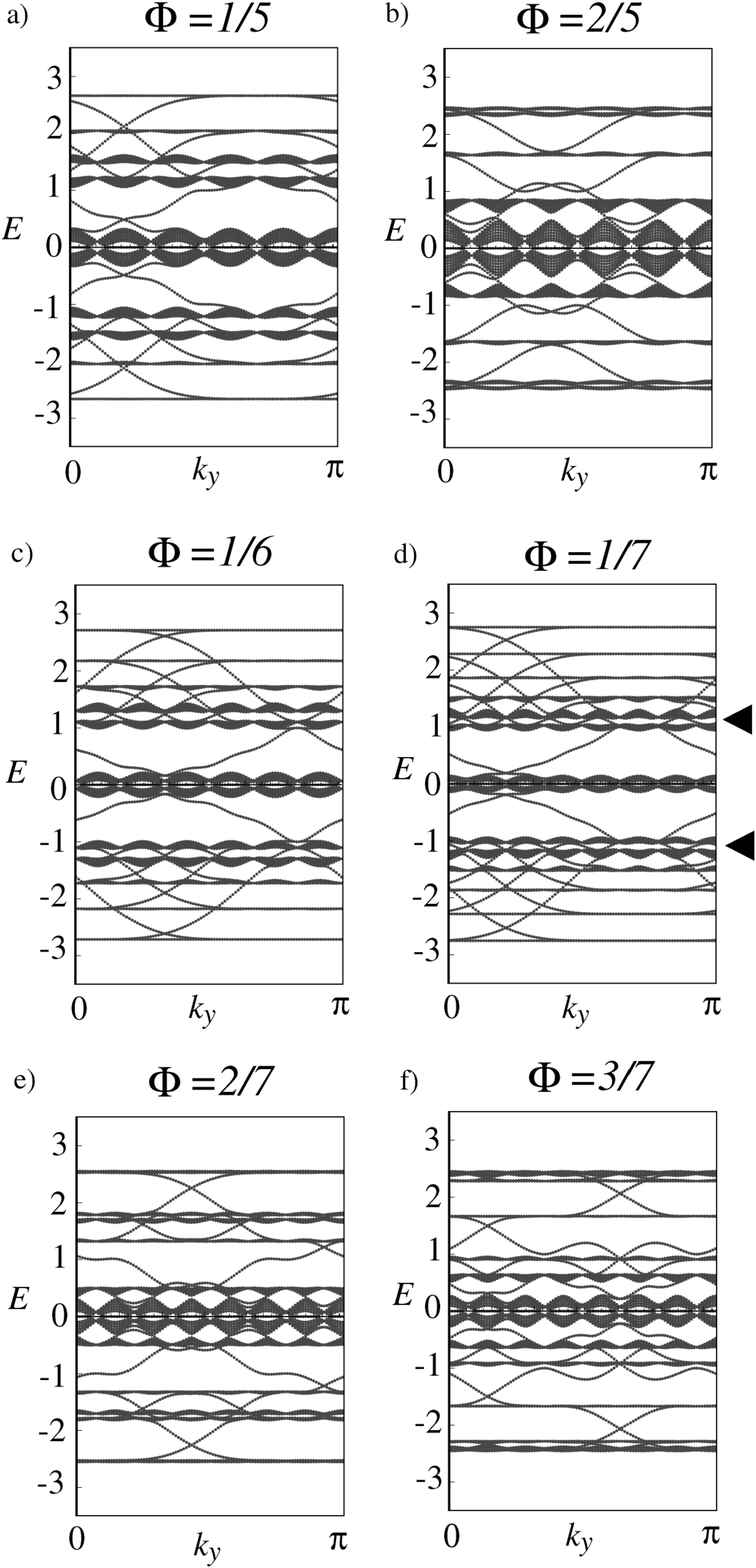}
\caption{
The energy spectra of graphene (honeycomb lattice with $t_a=1$) with
 bearded edges for a) $\Phi=1/5$, b) $\Phi=2/5$, c) $\Phi=1/6$, d)
 $\Phi=1/7$, e) $\Phi=2/7$ and d) $\Phi=3/7$. Here $k_y$ denotes the
 momentum along the edges. Edge states are seen in the energy gaps in
 the spectra. The arrows in d) indicate
 gaps where the number of edge states disagrees with the Hall
 conductance obtained by (\ref{eq:rule1}) and (\ref{eq:rule2}). Except
 these gaps, we have agreements between the numbers of edge states and the
 bulk Hall conductances determined by (\ref{eq:rule1}) and (\ref{eq:rule2}).  
}
\label{fig:edgestate}
\end{center}
\end{figure}

In Fig.\ref{fig:edgestate}, we illustrate the energy spectra for
$5\le q \le 7$. 
Except for gaps near the singularities mentioned above, 
we have excellent agreements between the number of the edge states and the bulk
Hall conductance determined by the rule (\ref{eq:rule1}) and
(\ref{eq:rule2}).
In particular, even for the subband gaps, which become invisible in a
weak magnetic field, these two quantum numbers are found to agree with
each other.
This results clearly indicate that except for gaps near the
singularities, the Hall conductances in graphene including those for subband gaps are determined by the rule (\ref{eq:rule1}) and
(\ref{eq:rule2}).
In addition,
for some gaps located just beyond the van Hove singularity at $E=1$ or
those on the lower and upper edges of the spectra, we find a disagreement
between them, which is also consistent with the analyses presented in the
previous section.
We also checked that the similar results hold for all $\Phi=p/q$ with
$q\le 11$.
We summarize the results for the bulk Hall conductance and the edge Hall
conductance in Tables \ref{table:table1} and \ref{table:table2}.

\begin{figure}[ht]
\begin{center}
\includegraphics[width=7cm]{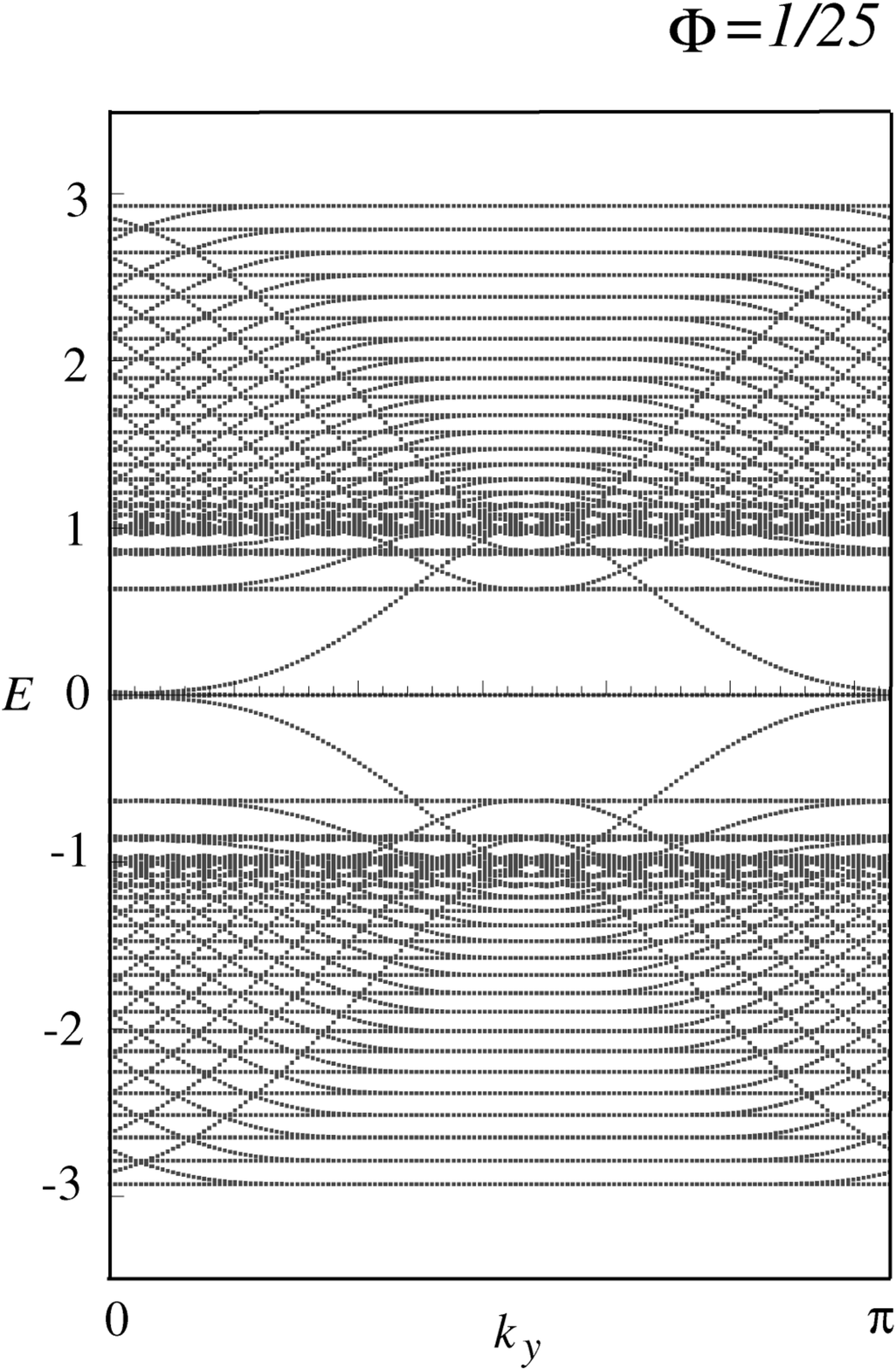}
\caption{
The energy spectrum of graphene with bearded edges for $\Phi=1/25$.
Here $k_y$ is the momentum along the edges. By counting edge states,
 we have $\sigma^{\rm edge}_{xy}=-(e^2/h)(2n+1)$ with $n=-2,-1,0,1$ near
 $E=0$.
 For the other regions, we have the conventional quantization of the Hall
 conductance.}
\label{fig:edge_p1q25}
\end{center}
\end{figure}

To illustrate a weak magnetic field case, the energy spectrum with
bearded edges for $\Phi=1/25$ is also shown in Fig.\ref{fig:edge_p1q25}. 
In a weak magnetic field, subband gaps are very tiny so only the edge
states for visible gaps are shown in the figure.
The unconventional quantization for the edge quantum Hall conductance is found
to persist up the van Hove singularities at $E=\pm 1$. 
We also find that for gaps beyond the van Hove singularities
the conventional quantization for the edge quantum Hall conductance holds.
The latter result suggests that 
the rule (\ref{eq:rule1}) and (\ref{eq:rule2}) is no
longer valid for some visible gaps located just beyond the van Have
singularities.
These results are also consistent with those in the previous section.

\begin{table}
\begin{center}
\begin{tabular}[t]{|c|c|c|c|}
\multicolumn{4}{l}{a) $\Phi=1/5$} \\ 
 \hline
$r$ &$(s_r^0,t_r^0)$& $\sigma_{xy}^{\rm weak}$ & $\sigma_{xy}^{\rm edge}$\\ 
\hline 
1 & (0,1) &1 &1 \\ 
2 & (0,2) & 2 &2 \\
3 & (1,-2) & 3 &3 \\
4 & (1,-1) & -1 &-1 \\
\hline
\end{tabular} 
\hspace{6ex}
\begin{tabular}[t]{|c|c|c|c|}
\multicolumn{4}{l}{b) $\Phi=2/5$} \\ 
 \hline
$r$ &$(s_r^0,t_r^0)$& $\sigma_{xy}^{\rm weak}$ & $\sigma_{xy}^{\rm edge}$\\ 
\hline 
1 & (1,-2) &3 &3 \\ 
2 & (0,1) & 1 &1 \\
3 & (1,-1) & -1 &-1 \\
4 & (0,2) & 2 &2\\
\hline
\end{tabular} 
\hspace{6ex}
\begin{tabular}[t]{|c|c|c|c|}
\multicolumn{4}{l}{c) $\Phi=1/6$} \\ 
 \hline
$r$ &$(s_r^0,t_r^0)$& $\sigma_{xy}^{\rm weak}$ & $\sigma_{xy}^{\rm edge}$\\ 
\hline 
1 & (0,1) &1 &1 \\ 
2 & (0,2) & 2 &2 \\
3 & (0,3) or (1,-3) & 3 or -3 & 3 \\
4 & (1,-2) & 4 &4 \\
5 &  (1,-1) &-1 & -1 \\
\hline
\end{tabular} 
\\
\vspace{5ex}
\begin{tabular}[t]{|c|c|c|c|}
\multicolumn{4}{l}{d) $\Phi=1/7$} \\ 
 \hline
$r$ &$(s_r^0,t_r^0)$& $\sigma_{xy}^{\rm weak}$ & $\sigma_{xy}^{\rm edge}$\\ 
\hline 
1 & (0,1) &1 &1 \\ 
2 & (0,2) & 2 &2 \\
3 & (0,3) & 3 &3 \\
4 & (1,-3) & -3 & 4 \\
5 & (1,-2) & 5 & 5 \\
6 & (1,-1) & 1 &1\\
\hline
\end{tabular} 
\hspace{6ex}
\begin{tabular}[t]{|c|c|c|c|}
\multicolumn{4}{l}{e) $\Phi=2/7$} \\ 
 \hline
$r$ &$(s_r^0,t_r^0)$& $\sigma_{xy}^{\rm weak}$ & $\sigma_{xy}^{\rm edge}$\\ 
\hline 
1 & (1,-3) &-3 &-3 \\ 
2 & (0,1) & 1 &1 \\
3 & (1,-2) & 5 &5 \\
4 & (0,2) & 2 & 2 \\
5 & (1,-1) &-1  & -1 \\
6 & (0,3) & 3 &3\\
\hline
\end{tabular} 
\hspace{6ex}
\begin{tabular}[t]{|c|c|c|c|}
\multicolumn{4}{l}{f) $\Phi=3/7$} \\ 
 \hline
$r$ &$(s_r^0,t_r^0)$& $\sigma_{xy}^{\rm weak}$ & $\sigma_{xy}^{\rm edge}$\\ 
\hline 
1 & (1,-2) &5 &5 \\ 
2 & (-1,3) & 3 &3 \\
3 & (0,1) & 1 &1 \\
4 & (1,-1) & -1 & -1 \\
5 & (2,-3) & -3 & -3 \\
6 & (0,2) & 2 &2\\
\hline
\end{tabular} 
\end{center} 
\caption{The bulk Hall conductance $\sigma_{xy}^{\rm weak}$ 
(in units of $-(e^2/h)$) of $r$-th gap from the bottom 
determined by (\ref{eq:rule1}) and (\ref{eq:rule2}), and the
 corresponding edge Hall conductance $\sigma_{xy}^{\rm edge}$
 determined by counting the edge states  for a) $\Phi=1/5$, b)
 $\Phi=2/5$, c) $\Phi=1/6$, d) $\Phi=1/7$, e) $\Phi=2/7$ and f)
 $\Phi=3/7$. The edge states are counted for $t_a=1$.
We also show the solution of the Diophantine equation $(s_r^0,t_r^0)$
 satisfying $|t_r^0|\le q/2$.
Here we have shown $\sigma_{xy}^{\rm weak}$ and $\sigma_{xy}^{\rm edge}$
for gaps in the lower band ($E<0$) since those for the $(2q-r)$th gap
 in the upper band are the same as those for $r$th
 gap in the lower band with opposite signs.} 
\label{table:table1}
\end{table}

\begin{table}
\begin{center}
\begin{tabular}{|c|c|c|c|c|}
\hline
 $\Phi$ &$r$&$(s_r^0,t_r^0)$
 &$\sigma_{xy}^{\rm weak}$ &$\sigma_{xy}^{\rm  edge}$ \\
\hline
1/7 & 4 & (1,-3) & 3 & -4 \\ 
    & 10 & (1,3) & -3 & 4  \\
\hline
1/8 & 5 & (1,-3) & 3 & -5 \\
    & 11  & (1,3) & -3 & 5 \\  
\hline
1/9 & 6 & (1,-3) & 3 & -6 \\
    & 12 & (1,3) & -3 & 6 \\
\hline 
4/9 & 2 & (2,-4) & 4 & -5 \\
    & 16 & (0,4) & -4 & 5 \\
\hline
1/11 & 6 & (1,-5) & 5 & -6 \\
     & 16 & (1,5) & -5 & 6 \\
\hline
5/11 & 2 & (2,-4) & 4 & -7 \\
     & 20 & (0,4) & -4 & 7 \\
 \hline 
\end{tabular}
\end{center} 
\caption{A list of gaps where $\sigma_{xy}^{\rm edge}$
 disagrees with $\sigma_{xy}^{\rm weeak}$ $(q\le 11)$ .
The edge states are counted for $t_a=1$. For $\Phi=4/9$ and $\Phi=5/11$,
 the gaps are located on the lower and upper edges of the spectra, and for the
 other cases, they are located just beyond the van Hove singularity.
 All of them are located at $|E|>1$.
 }
\label{table:table2}
\end{table}

\section{Conclusion}
\label{sec:conclusion}
In this paper, we examined quantum phase structures of a tight-binding
model on the honeycomb lattice, which are controlled by the
hopping parameter $t_a$.
In contrast to the square lattice, where no phase transition occurs by
changing the ratio of hopping parameters $t_x/t_y$,
it was found that phase transitions occur as we change $t_a$. 
In terms of the Diophantine equation, we characterized the weak
$t_a $ phase and the strong $t_a$ one, respectively, and determined the Hall
conductances for both cases. 
We found that 
the weak $t_a$ phase shows the unconventional quantization of the Hall
conductance in a weak magnetic field although the strong $t_a$ phase
shows only the conventional one.   
This implies the existence of quantum phase transitions accompanying
gap closing points in the intermediate $t_a$ region, which
was confirmed by numerical calculations. 
We also found numerically that unexpected quantum phase transitions occur
for some gaps just beyond the van Hove singularities at $t_a=1$ or those
on the upper and lower edges of the spectrum.

Using these results, 
we analyzed in detail the Hall conductance in graphene ($t_a=1$).
Except for some gaps just beyond the van Hove singularities or on the edges
of the spectrum, the Hall conductances including those for the subband
gaps were determined.
They naturally explain the experimentally observed unconventional
quantization of the quantum Hall effect in a weak magnetic field.
They also predict that the unconventional quantization persists up to the
van Hove singularity. 
We also examined the edge states in graphene and confirmed the bulk-edge
correspondence in quantum Hall effect.

\newpage 
\bibliography{topological_order}

\end{document}